\documentclass[letterpaper,journal]{IEEEtran}
\usepackage{amsmath,amsfonts}
\usepackage{algorithmic}
\usepackage{algorithm}
\usepackage{array}
\usepackage[caption=false,font=normalsize,labelfont=sf,textfont=sf]{subfig}
\usepackage{textcomp}
\usepackage{stfloats}
\usepackage{url}
\usepackage{verbatim}
\usepackage{graphicx}
\usepackage{cite}
\usepackage[table,xcdraw]{xcolor} 
\usepackage{xcolor}
\usepackage{adjustbox}
\usepackage{hyperref}

\usepackage{colortbl}
\usepackage{tikz}
\usetikzlibrary{arrows.meta, positioning, shadows, backgrounds, shadings}
\usepackage{booktabs}
\usepackage{enumitem}
\usepackage{multirow}
\newcolumntype{L}[1]{>{\raggedright\let\newline\\\arraybackslash\hspace{0pt}}p{#1}}
\usepackage{ragged2e} 
\usepackage{makecell}

\let\oldcite\cite
\renewcommand{\cite}[1]{\mbox{\oldcite{#1}}}

\definecolor{revblueII}{RGB}{0,0,255}
\newcommand{\rvnew}[1]{\textcolor{revblueII}{#1}}

\colorlet{revblue}{black}
\colorlet{revblueII}{black}

\begin{document}


\title{A Survey on Medical Image Compression: From Traditional to Learning-Based Approaches\thanks{This work did not involve human subjects or animals in its research.}}


\author{
Guofeng Tong,~\IEEEmembership{}%
\thanks{Fenglei Fan is the corresponding author. (e-mail: fenglfan@cityu.edu.hk and hitfanfenglei@gmail.com).}%
\thanks{Guofeng Tong is with School of Cyberspace Science and Technology, Beijing Jiaotong University, Beijing, China. (e-mail: 22120506@bjtu.edu.cn).}%
\and
Sixuan Liu,~\IEEEmembership{}%
\thanks{Sixuan Liu is with Institute for Brain and Cognitive Sciences, Department of Automation, Tsinghua University, Beijing, China. (e-mail: 13283012@bjtu.edu.cn).}%
\and
Yang Lv,~\IEEEmembership{}%
\thanks{Yang Lv is with Molecular Imaging Business Unit, Shanghai United Imaging Healthcare Co., Ltd, Shanghai, China. (e-mail: yang.lv@united-imaging.com).}%
\and
Hanyu Pei, 
Feng-Lei Fan~\IEEEmembership{}%
\thanks{Hanyu Pei and Feng-Lei Fan are with Frontier of Artificial Network (FAN) Lab, Department of Data Science, City University of Hong Kong, Hong Kong, China SAR.}%
}

\markboth{submitted to IEEE Transactions on Radiation and Plasma Medical Sciences, Vol. XX, No. XX, XX 2024}%
{Tong \MakeLowercase{\textit{et al.}}: A Survey on Medical Image Compression}

\IEEEpubid{0000--0000/00\$00.00~\copyright~2021 IEEE}

\maketitle

\begin{abstract}

The exponential growth of medical imaging has created significant challenges in data storage, transmission, and management for healthcare systems. In this vein, efficient compression becomes increasingly important. Unlike natural image compression, medical image compression prioritizes preserving diagnostic details and structural integrity, imposing stricter quality requirements and demanding fast, memory-efficient algorithms that balance computational complexity with clinically acceptable reconstruction quality. Meanwhile, the medical imaging family includes a plethora of modalities, each possessing different requirements. For example, compression of 2D medical images (e.g., X-rays, histopathological images) focuses on exploiting intra-slice spatial redundancy, whereas 3D volumetric medical images require modeling both intra-slice and inter-slice spatial correlations, and 4D dynamic volumetric imaging (e.g., time-series CT/MRI and 4D ultrasound) additionally demands capturing temporal correlations between consecutive frames. Traditional compression methods, grounded in mathematical transforms and information theory principles, provide solid theoretical foundations, predictable performance, and high standardization levels, with extensive validation in clinical environments. In contrast, deep learning-based approaches demonstrate remarkable adaptive learning capabilities and can capture complex statistical characteristics and semantic information within medical images. This comprehensive survey establishes a two-facet taxonomy based on data structure (2D vs 3D/4D) and technical approaches (traditional vs learning-based), thereby systematically presenting the complete technological evolution, analyzing the unique technical challenges, and prospecting future directions in medical image compression.

\end{abstract}


\section{Introduction}
\IEEEPARstart{T}{he} rapid advancement of medical imaging technologies has fundamentally shaped modern healthcare, enabling unprecedented diagnostic capabilities and therapeutic interventions. Since Wilhelm Conrad Röntgen's groundbreaking discovery of X-rays in 1895~\cite{rontgen1895}, the field has evolved from the basic radiographic imaging to sophisticated modalities such as Magnetic Resonance Imaging (MRI), Computed Tomography (CT), ultrasound, and nuclear medicine imaging~\cite{seibert1995}. The introduction of CT scanning by Godfrey Hounsfield and Allan Cormack in the 1970s~\cite{hounsfield1973}, followed by the development of the MRI technology by Paul Lauterbur and Peter Mansfield in the same decade~\cite{lauterbur1973}, has established medical imaging as an indispensable component of contemporary clinical practice. These technologies generate a vast amount of high-resolution, high-fidelity images that serve as a critical diagnostic and prognostic tool for patients.\IEEEpubidadjcol

However, the widespread adoption of medical imaging presents significant challenges for healthcare systems~\cite{langer2011challenges}. Modern medical imaging equipment routinely produces images with unprecedented spatial resolution, temporal sampling rates, and bit depths. For instance, a single high-resolution CT examination can generate hundreds of cross-sectional images~\cite{wang2018improving}, while dynamic MRI studies produce thousands of volumetric datasets over time~\cite{slavin2005spatial}. The storage requirement for medical images has reached substantial proportions, with even individual studies often taking gigabytes of storage space. When multiplied across the entire hospital serving tens of thousands of patients annually, the cumulative storage and transmission requirements are hard to withstand, necessitating efficient data compression strategies~\cite{dicom_standard,skodras2001jpeg}. These medical images can be broadly categorized into two major types: 2D images and volumetric data, including static 3D volumes and 4D dynamic volumetric imaging. Typical 2D images include X-ray radiographs and histopathological images, whereas volumetric data include multi-slice CT sequences and 3D MRI datasets; 4D dynamic volumetric imaging further extends these volumes over time.

\begin{figure}[H]
	\centering
	\includegraphics [width=\linewidth]{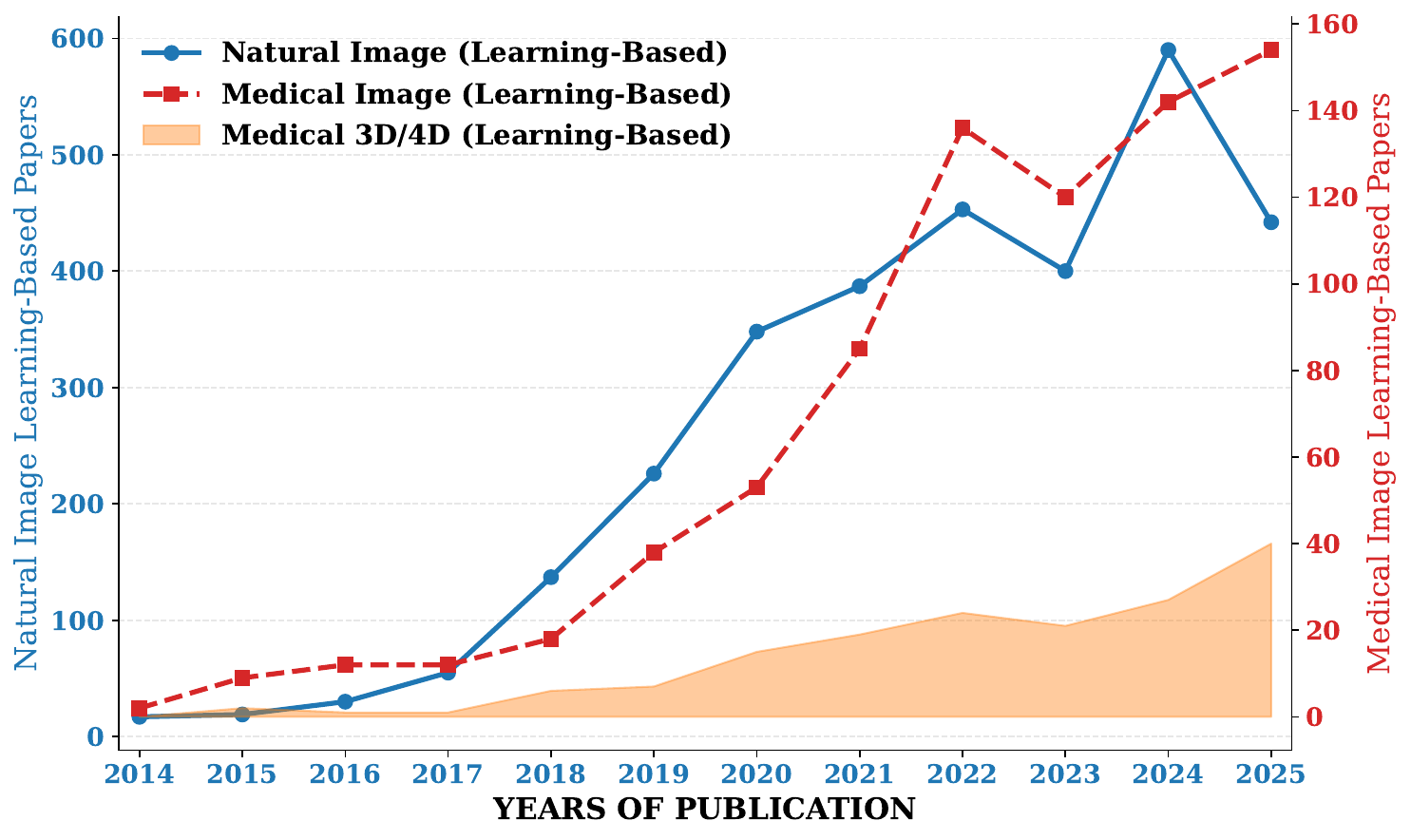}\\
	\caption{Evolution of Learning-Based Compression: Natural vs. Medical. Note that a dual y-axis is used: the left axis corresponds to natural-image papers (0–600), while the right axis corresponds to medical-image papers (0–160). Therefore, curve heights are not directly comparable across domains.}
	\label{fig:med_vs_natural_trend}
    \vspace{0cm}
\end{figure}

Preserving diagnostic information while achieving efficient storage and transmission has driven extensive research in medical image compression~\cite{me2012survey,kalaiselvi2021significant}. We analyze the publication trend using data retrieved from the Web of Science Core Collection from 2014 to 2025, as illustrated in Figure~\ref{fig:med_vs_natural_trend}. Specifically, three structured Boolean search queries were used to distinguish (1) learning-based natural image compression, (2) learning-based medical image compression, and (3) learning-based medical \mbox{3D/4D} compression.
For natural images, we searched articles using:
\texttt{("Image Compression" OR "Image Coding" OR "Video Compression" OR "Video Coding") AND ("Deep Learning" OR "Neural Network" OR "CNN" OR "Transformer" OR "Generative" OR "Autoencoder") NOT ("Medical" OR "CT" OR "MRI" OR "Ultrasound" OR "X-ray")}. For medical images, we applied:
\texttt{(("Medical Image" OR "CT" OR "MRI" OR "Ultrasound" OR "X-ray") AND ("Compression" OR "Coding") AND ("Deep Learning" OR "Neural Network" OR "CNN" OR "Transformer" OR "Generative"))}. For medical \mbox{3D/4D}, we extended the medical query by appending the keywords: \texttt{("3D" OR "Volumetric" OR "4D" OR "Dynamic")}. The dual-axis chart has two highlights: First, in recent years, the number of publications on medical image compression shows a rapid increase, which reflects this domain has attracted increasing attention. Second, Figure~\ref{fig:med_vs_natural_trend} reveals a clear correlation between the natural and medical imaging domains. A distinct technological lag is observable, where the rapid growth in learning-based natural image compression predates that in the medical domain by approximately 2--3 years. Nowadays, learning-based approaches are more and more adopted in medical imaging. Most notably, volumetric methods are rapidly gaining traction, specifically roughly doubling in publication volume from 2023 to 2025, which confirms that leveraging inter-slice and temporal correlations is the current frontier in medical image compression research.
\color{black}

However, medical image compression operates under fundamentally different constraints compared to general-purpose image compression, which necessitates specialized approaches. To systematically analyze the technical challenges and development trends in medical image compression, we establish an analytical framework based on three fundamental dimensions that constitute the core organizational principles of this survey:
\begin{itemize}
    \item Application domain differences - The essential distinctions between medical image compression and natural image compression

\item Data structure differences - The compression challenges of 2D images versus 3D/4D medical images\IEEEpubidadjcol

\item Technical approach differences - The comparative advantages of traditional mathematical methods versus learning-based approaches
\end{itemize}
The intersection of these three dimensions is the core of medical image compression research, providing a systematic framework for understanding the applicability and development directions of different methods.

\begin{figure}[!t]
	\centering
	\includegraphics [width=\linewidth]{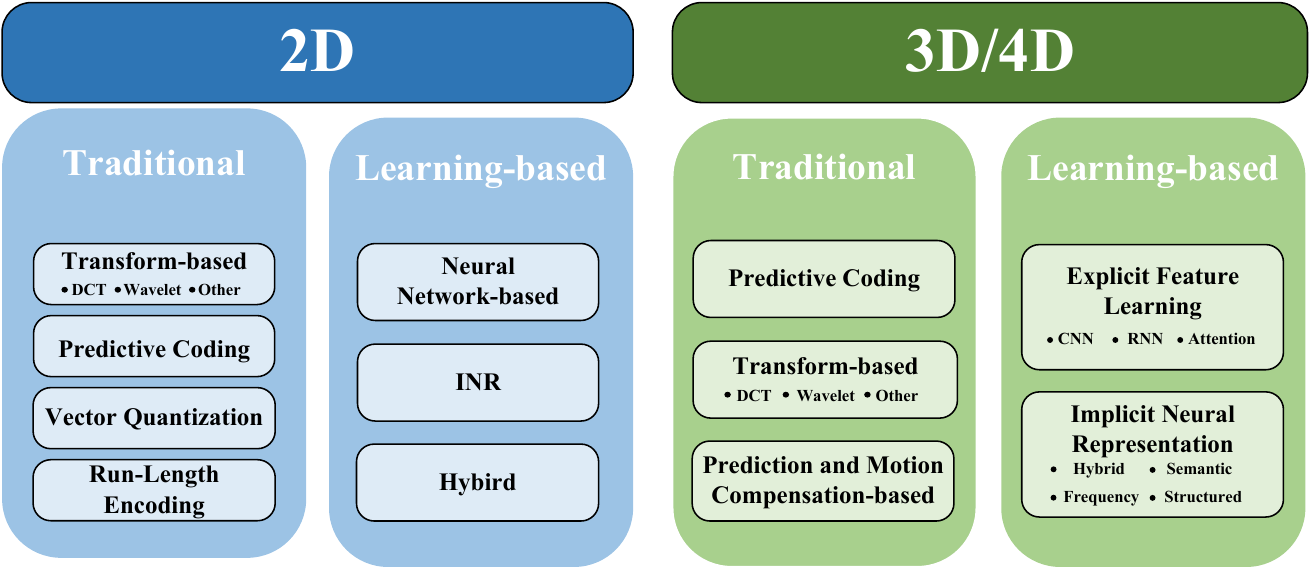}\\
	\caption{Taxonomy used for this medical image compression review.}
	\label{fig:taxonomy}
    \vspace{-0.5cm}
\end{figure}

\definecolor{headerbg}{RGB}{230, 236, 240} 
\definecolor{rowodd}{RGB}{255, 255, 255}   
\definecolor{roweven}{RGB}{245, 248, 250}  
\definecolor{mywork}{RGB}{255, 246, 235}   

\begin{table*}[t]
\centering
\caption{Comparison of Related Survey Articles on Image and Medical Image Compression}
\label{tab:survey_comparison}

\setlength{\aboverulesep}{0pt}
\setlength{\belowrulesep}{0pt}
\renewcommand{\arraystretch}{1.5} 
\setlength{\tabcolsep}{5pt}       

\begin{tabular}{l c c c c c c m{7.5cm}}
\toprule

\rowcolor{headerbg}
\textbf{Survey} & \textbf{Year} & \multicolumn{5}{c}{\textbf{Scope}} & \textbf{Contributions and Limitations} \\
\cmidrule(lr){3-7}
\rowcolor{headerbg}
 &  
& \textbf{2D} & \textbf{3D/4D} & \textbf{Trad.} 
& \textbf{DL} & \textbf{INR} 
&  \\ 
\midrule


\rowcolor{rowodd}
\textbf{Ma et al.\cite{ma2019image}} & 2019 
& \checkmark & -- & -- & \checkmark & -- 
& \scalebox{0.7}{$\bullet$} Comprehensive review of neural-network-based image/video compression. \newline
  \scalebox{0.7}{$\bullet$} Covers CNN/RNN frameworks. \newline
  $\textit{Limitation:}$ General-purpose; no medical imaging focus. \\

\rowcolor{roweven}
\textbf{Mishra et al.\cite{mishra2022deep}} & 2022 
& \checkmark & -- & -- & \checkmark & -- 
& \scalebox{0.7}{$\bullet$} Critical review of deep architectures (AE, CNN, VAE, GAN). \newline
  \scalebox{0.7}{$\bullet$} Summarizes benchmarks and RD metrics. \newline
  $\textit{Limitation:}$ Not tailored to medical imaging constraints. \\

\rowcolor{rowodd}
\textbf{Boopathiraja et al.\cite{boopathiraja2022computational}} & 2022 
& \checkmark & \checkmark & \checkmark & Partial 
& -- 
& \scalebox{0.7}{$\bullet$} Survey of computational 2D/3D medical compression. \newline
  \scalebox{0.7}{$\bullet$} Covers predictive, lossless, and early ML methods. \newline
  $\textit{Limitation:}$ Lacks modern generative models/INR/4D. \\

\rowcolor{roweven}
\textbf{Molaei et al.\cite{molaei2023implicit}} & 2023 
& Partial & Partial & -- & Partial 
& \checkmark 
& \scalebox{0.7}{$\bullet$} First survey on INR in medical imaging. \newline
  \scalebox{0.7}{$\bullet$} Covers reconstruction, SR, and compression. \newline
  $\textit{Limitation:}$ Lacks comparison with traditional codecs. \\

\rowcolor{rowodd}
\textbf{Bourai et al.\cite{bourai2024deep}} & 2024 
& \checkmark & Partial & -- & \checkmark 
& -- 
& \scalebox{0.7}{$\bullet$} Review of DL-assisted medical compression. \newline
  \scalebox{0.7}{$\bullet$} Analyzes diagnostic fidelity and clinical constraints. \newline
  $\textit{Limitation:}$ Limited traditional methods and 4D data. \\

\rowcolor{mywork} 
\textbf{This Survey} & \textbf{2025} 
& \textbf{\checkmark} & \textbf{\checkmark} & \textbf{\checkmark} & \textbf{\checkmark} & \textbf{\checkmark} 
& \scalebox{0.7}{$\bullet$} \textbf{Unified Taxonomy:} Integrates classical, deep neural, and INR methods across 2D, 3D, and 4D. \newline
  \scalebox{0.7}{$\bullet$} \textbf{Clinical Focus:} Discusses DICOM standards, diagnostic quality, and workflow integration. \\
\bottomrule

\end{tabular}
\end{table*}

Based on the above framework, we propose a two-axis taxonomy for medical image compression (Fig.~\ref{fig:taxonomy}), organized by data structure (2D vs.~3D/4D) and methodology (traditional vs.~learning-based). This yields four categories with distinct characteristics and application scenarios. We use this taxonomy to structure the survey and to formulate three guiding questions.

\textbf{Q1. How does medical image compression differ from natural image compression?}
Medical image compression is constrained by stricter diagnostic-fidelity requirements, regulatory compliance, and clinical workflow needs. Unlike natural images where moderate perceptual distortion may be acceptable, medical images often require lossless or near-lossless fidelity depending on modality and task. When applicable, clinical systems may also benefit from ROI-aware coding to treat diagnostically critical regions differently from background.

\rvnew{
Moreover, medical images are commonly stored in DICOM with higher bit-depth and modality-dependent intensity scales (e.g., physically calibrated CT values), leading to different statistics and error tolerances than 8-bit RGB images. Many acquisitions are volumetric or dynamic (3D/4D), where inter-slice and temporal correlations provide additional redundancy exploitable by 3D/4D prediction or context modeling. Finally, anatomical structure priors motivate edge/structure-aware transforms and ROI-adaptive bit allocation.}

\textit{\textbf{Q2}. \textbf{What fundamental differences exist between 2D and \mbox{3D/4D} volumetric medical image compression}?}
2D medical image compression can leverage mature 2D transform and predictive coding techniques to exploit intra-slice spatial redundancy while meeting diagnostic-quality requirements. In contrast, volumetric medical images pose additional challenges: static 3D data require modeling both intra-slice and inter-slice correlations, while dynamic 4D imaging further demands capturing temporal correlations across consecutive frames. Compressing volumes slice-by-slice with 2D codecs ignores these multi-dimensional dependencies and typically leads to suboptimal coding efficiency~\cite{xiong2003lossy}.

The complexity of volumetric data arises from (i) strong inter-slice and temporal correlations that call for multi-dimensional spatial/spatio-temporal modeling, (ii) anisotropic sampling (e.g., slice spacing versus in-plane resolution and temporal sampling), and (iii) substantially increased memory and computational costs with higher dimensionality, which challenges efficiency and scalability~\cite{senapati2016volumetric,bruylants2015wavelet}. Therefore, volumetric medical image compression is not merely a straightforward extension of 2D methods but often requires dedicated multi-dimensional frameworks for 3D spatial and 4D spatio-temporal coding.

\textit{\textbf{Q3}. \textbf{How do traditional compression methods compare with learning-based approaches in medical image compression?}}
Traditional compression methods, grounded in mathematical transforms and information theory, offer solid theoretical foundations, predictable behavior, and high standardization, and are widely adopted in clinical systems~\cite{huang2011pacs}. Their deterministic designs facilitate quality assurance and compliance in regulated settings. In contrast, learning-based approaches can capture complex statistical dependencies (and, in some cases, higher-level semantics) in medical images, and have shown competitive or improved rate--distortion performance in reported studies~\cite{min2022lossless}. However, their data dependence, computational demands, and limited interpretability can complicate clinical deployment in terms of validation, regulatory clearance, and infrastructure integration~\cite{topol2019high}. Hybrid approaches aim to combine the reliability of traditional methods with the adaptability of learning-based models, and represent an important direction for medical image compression~\cite{paul2022,li2025towards}.

\begin{figure*}[!t]
\centering

\tikzset{
    bead/.style={
        circle, 
        shading=ball, 
        inner sep=2.5pt, 
        draw=white,      
        thick,
        circular drop shadow={shadow scale=1.05, shadow xshift=0.5pt, shadow yshift=-0.5pt, opacity=0.3}
    },
    style2DTrad/.style={bead, ball color=gray!60},
    style3DTrad/.style={bead, ball color=cyan!50},
    style2DLB/.style={bead, ball color=blue!70!black},
    style3DLB/.style={bead, ball color=red!70!orange},
    labeltext/.style={font=\sffamily\scriptsize, align=center},
    axislabel/.style={font=\sffamily\footnotesize\bfseries, fill=white, inner sep=2pt}
}

\begin{tikzpicture}[
    >=Stealth, 
    x=0.32cm, 
    y=1.2cm   
]

\begin{scope}[on background layer]
    \fill[gray!5] (0,-0.5) rectangle (14.5, 4.2);
    \node[anchor=south, gray!60, font=\sffamily\scriptsize\bfseries] at (7.25, 3.8) {PHASE I: 1970--2000};
    
    \fill[blue!3] (14.5,-0.5) rectangle (31.5, 4.2);
    \node[anchor=south, blue!40!gray, font=\sffamily\scriptsize\bfseries] at (23, 3.8) {PHASE II: 2000--2015};

    \fill[orange!5] (31.5,-0.5) rectangle (48,-0.5 |- 0,4.2);
    \node[anchor=south, orange!50!gray, font=\sffamily\scriptsize\bfseries] at (38, 3.8) {PHASE III: 2015--2025};
\end{scope}

\foreach \y in {0,1,2,3} \draw[gray!20, line width=0.8pt] (0,\y) -- (48,\y);

\draw[->, thick, gray!80] (0,-0.1) -- (48.5,-0.1) node[below left, text=black] {\footnotesize Year};

\foreach \x/\year in {0/1970, 14/2000, 31/2015, 45/2025} {
    \draw[gray, thick] (\x, -0.1) -- (\x, -0.25);
    \node[below, font=\sffamily\footnotesize] at (\x, -0.25) {\year};
}

\node[anchor=east, font=\sffamily\scriptsize\bfseries, align=right, color=gray!70] at (-0.5,3) {2D\\Traditional};
\node[anchor=east, font=\sffamily\scriptsize\bfseries, align=right, color=cyan!60!black] at (-0.5,2) {3D/4D\\Traditional};
\node[anchor=east, font=\sffamily\scriptsize\bfseries, align=right, color=blue!70!black] at (-0.5,1) {2D\\Deep Learning};
\node[anchor=east, font=\sffamily\scriptsize\bfseries, align=right, color=red!70!orange] at (-0.5,0) {3D/4D\\Deep Learning};

\node[style2DTrad] (dct74) at (2,3) {};
\node[labeltext, above=2pt of dct74] {DCT~[\cite{DCT1974}]};

\node[style2DTrad] (wav94) at (12,3) {};
\node[labeltext, above=2pt of wav94] {2D Wavelet~[\cite{lewis1992image}]};

\node[style2DTrad] (jp2k02) at (17,3) {};
\node[labeltext, below=2pt of jp2k02] {JPEG2000~[\cite{skodras2001jpeg}]};

\node[style3DTrad] (volwav92) at (10,2) {};
\node[labeltext, above=2pt of volwav92] {Vol.~Wavelet~[\cite{muraki1992approximation}]};

\node[style3DTrad] (loss3d99) at (14,2) {};
\node[labeltext, below=2pt of loss3d99] {Lossless 3D~[\cite{kim1999lossless}]};

\node[style3DTrad] (jp3d09) at (22,2) {};
\node[labeltext, above=2pt of jp3d09] {JP3D~[\cite{bruylants2009jp3d}]};

\node[style2DLB] (mxid21) at (34,1) {};
\node[labeltext, above=2pt of mxid21] {Med.~CAE~[\cite{fettah2024convolutional}]};

\node[style2DLB] (vae22) at (38,1) {};
\node[labeltext, below=2pt of vae22] {Med.~VAE~[\cite{liu2022medical}]};

\node[style2DLB] (hyb22) at (42,1) {};
\node[labeltext, above=2pt of hyb22] {Hybrid~[\cite{paul2022}]};

\node[style3DLB] (cnn3d22) at (34,0) {};
\node[labeltext, above=2pt of cnn3d22] {3D CNN~[\cite{luu2021efficiently}]};

\node[style3DLB] (rnn22) at (37,0) {};
\node[labeltext, below=2pt of rnn22] {RNN/LSTM~[\cite{nagoor2021medzip}]};

\node[style3DLB] (sinco23) at (43,0) {};
\node[labeltext, above=2pt of sinco23] {INR (SINCO)~[\cite{gao2023sinco}]};

\begin{scope}[shift={(10, 4.8)}] 
    \node[style2DTrad] (L1) at (0,0) {}; 
    \node[right=2pt of L1, font=\sffamily\scriptsize] {2D Trad.};
    
    \node[style3DTrad] (L2) at (8,0) {}; 
    \node[right=2pt of L2, font=\sffamily\scriptsize] {3D Trad.};
    
    \node[style2DLB] (L3) at (16,0) {}; 
    \node[right=2pt of L3, font=\sffamily\scriptsize] {2D Learning};
    
    \node[style3DLB] (L4) at (25,0) {}; 
    \node[right=2pt of L4, font=\sffamily\scriptsize] {3D Learning};
\end{scope}

\end{tikzpicture}
\caption{Timeline of representative medical image compression methods. Each dot corresponds to a representative cited work, rendered with color gradients to distinguish the methodological taxonomy. The timeline is visually divided into three phases (1970--2000, 2000--2015, and 2015--2025) to emphasize the transition from transform-based standards to volumetric deep learning and implicit neural representations.}
\label{fig:timeline_taxonomy}
\end{figure*}


\color{revblue}
Understanding recent technological developments is essential for identifying the limitations of existing compression approaches and the opportunities for medical imaging. Several recent surveys on general-purpose image compression (e.g., \cite{ma2019image,mishra2022deep}) provide valuable overviews, but they primarily focus on natural images and devote limited attention to medical-specific constraints such as diagnostic-fidelity requirements, compliance considerations, and modality-dependent characteristics. Surveys dedicated to medical image compression also offer important insights, yet each typically covers only a subset of the rapidly expanding landscape. For instance, \cite{boopathiraja2022computational} provides a unified review of classical 2D/3D methods but offers limited coverage of recent advances in deep, generative, and implicit neural representation (INR) based techniques. More recent surveys \cite{bourai2024deep,molaei2023implicit} focus respectively on deep neural compression and INR-based methods. While timely, they place less emphasis on the theoretical foundations of classical approaches and do not provide a unified comparative perspective bridging traditional, deep, generative, hybrid, and INR-based techniques.

Given the rapid evolution of learning-based and hybrid methods, we present a consolidated survey that connects modern developments with longstanding classical techniques and with practical clinical requirements. Our contributions are three-fold:

\begin{itemize}
\item \textit{Systematic taxonomy}: We propose a two-axis taxonomy by data structure (2D vs.~3D/4D) and methodology (traditional vs.~learning-based), and align it with a historical timeline of representative methods.
\item \textit{Comprehensive technical coverage}: We review the spectrum from classical transform/predictive coding to modern deep, generative, hybrid, and INR-based approaches.
\item \textit{Medical domain specialization}: We emphasize modality-specific characteristics and clinical constraints (e.g., diagnostic fidelity, standards, and workflow integration) that shape codec design in practice.
\end{itemize}

\rvnew{
To address the breadth of the literature, for each category we describe a representative end-to-end codec pipeline and summarize other works by highlighting key differences and advancements.}

To put our survey in perspective, Table~\ref{tab:survey_comparison} compares representative surveys published between 2015 and 2024 along several dimensions, including covered modalities (2D vs.~3D/4D), methodological scope (traditional, deep, generative, hybrid, and INR-based), and whether medical-specific constraints (e.g., diagnostic fidelity, standards, and clinical workflow) are explicitly discussed. To the best of our knowledge, few existing works jointly examine these methodological families within a unified comparative framework while simultaneously accounting for modality-dependent characteristics and clinical constraints.

Figure~\ref{fig:timeline_taxonomy} provides a historical overview of representative medical image compression methods over the past five decades. The timeline highlights key shifts from early transform/predictive and entropy coding to modern learning-based approaches, and illustrates how developments in 2D and \mbox{3D/4D} compression have progressed in parallel. Importantly, the taxonomy adopted in this survey---distinguishing 2D versus \mbox{3D/4D} data structures and traditional versus learning-based methods---is aligned with the evolution depicted in Figure~\ref{fig:timeline_taxonomy}, motivating the organization of the subsequent sections.

\section{2D Medical Image Compression Methods}
2D medical images are fundamental in modern imaging (e.g., radiography, ultrasound, and histopathology). Their compression must preserve diagnostic information while enabling efficient storage and transmission. This section reviews 2D medical image compression methods, grouped into traditional approaches (signal processing and information theory) and learning-based methods that learn compact representations from data. We first discuss traditional methods, followed by recent learning-based advances.

\subsection{Traditional Medical Image Compression Methods}

\subsubsection{Transform-based Methods}
\begin{figure*}[htbp]
	\centering
	\includegraphics [width=0.8\textwidth]{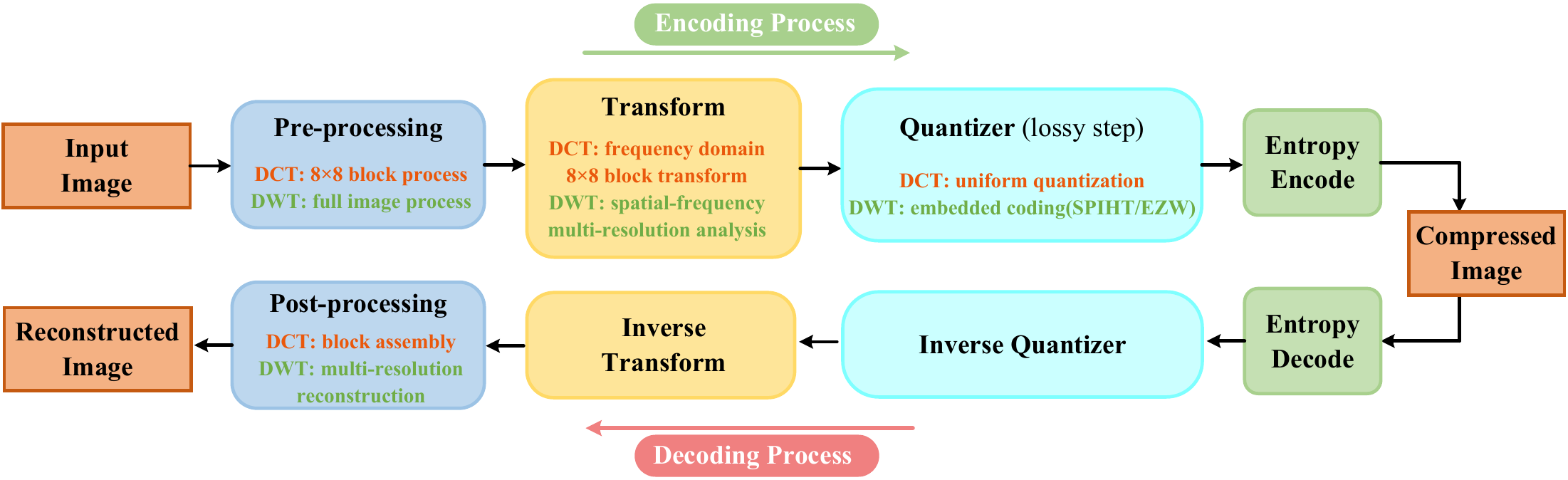}\\
	\caption{General framework of transform-based medical image compression methods.}
	\label{fig:2dtransform}
    \vspace{-0.5cm}
\end{figure*}

Transform-based methods are among the most established traditional techniques. They decorrelate spatial data via mathematical transforms, concentrating energy into a small number of significant coefficients, which can then be coded efficiently. \rvnew{A typical pipeline (Fig.~\ref{fig:2dtransform}) includes (i) partitioning/decomposition (e.g., block DCT or wavelet decomposition), (ii) transform, (iii) optional quantization for lossy coding, and (iv) entropy coding; decoding performs the inverse operations.} In lossless settings, reversible (integer-to-integer) transforms are typically used without irreversible quantization.

Transform-based compression relies on energy compaction: an effective transform concentrates most signal energy into a small number of coefficients~\cite{ahmed2006discrete}. In medical imaging, this can be advantageous because many modalities (modality-dependent) exhibit piecewise-smooth anatomical structures with prominent edges, which can be represented efficiently in suitable transform domains~\cite{mallat2002theory,suetens2017fundamentals}. Moreover, transform coding is closely related to rate--distortion optimization principles~\cite{cover1999elements}, and in practice it often achieves competitive rate--distortion performance compared with simple spatial-domain schemes under comparable settings.

\textbf{DCT Methods}.
The discrete cosine transform (DCT) traces back to the pioneering work of Ahmed et al.~\cite{ahmed2006discrete}, which introduced DCT as a real-valued alternative to the discrete Fourier transform and highlighted its strong energy-compaction property. Its practical impact was established through adoption in the JPEG standard, described by Wallace~\cite{wallace1991jpeg}, where block-based DCT forms a core component of a widely used still-image coding pipeline. \rvnew{In medical imaging, DCT-based codecs can be viewed as an instantiation of Fig.~\ref{fig:2dtransform} and typically use a block-based workflow consisting of block partitioning, DCT, optional thresholding or quantization, and entropy coding, with decoding performing the inverse operations.}

The evolution of DCT methods in medical image compression includes hybrid designs that modify the transform stage by combining DCT with other transforms. Singh et al.~\cite{singh2007dwt} introduced a DWT--DCT hybrid scheme that applies wavelet decomposition to obtain a multiresolution representation and then performs DCT on the resulting subbands. They reported improved compression performance under their evaluation settings compared with using either transform alone, while maintaining image quality suitable for medical interpretation.

Concurrently, Chen~\cite{chen2007medical} proposed a DCT-based subband decomposition combined with a modified Set Partitioning in Hierarchical Trees (SPIHT) organization~\cite{said1996}, adapting a hierarchical coding strategy originally developed for wavelet coefficients. The method applies 8$\times$8 DCT to form subbands and then encodes the coefficients using a modified hierarchical partitioning strategy. The authors reported high compression ratios, such as exceeding 20:1, while maintaining diagnostically acceptable quality on their reported datasets and settings. The design also supports progressive transmission through hierarchical coding.

Practical implementation and parameter selection were further studied by Chikouche et al.~\cite{chikouche2008application}, who combined DCT with arithmetic coding for MRI brain image compression. Their experiments examined thresholding and block-size choices and reported favorable trade-offs for thresholds in the range of 0--20 with block sizes of 16$\times$16 and 32$\times$32. Larger thresholds, such as above 30, led to pronounced quality degradation. These observations provide useful guidance for configuring DCT-based pipelines in medical settings, especially when balancing compression efficiency against artifact sensitivity.

Raj and Venkateswarlu~\cite{raj2007novel} further explored a sequential 3D-DCT strategy for medical image compression. Instead of applying DCT independently to each 2D block, their method groups a number of adjacent 2D pixel blocks into a 3D data cube and then performs a 3D DCT to decorrelate similar blocks before quantization and entropy coding. This design can be viewed as a block-level extension of the conventional DCT pipeline in Fig.~\ref{fig:2dtransform}, and is typically used in a slice-wise or single-image setting rather than exploiting true inter-slice volumetric correlations.

More recently, DCT-based medical image compression has been explored for specialized modalities. Yadav et al.~\cite{yadav2023near} investigated near-lossless compression of Tc-99m dimercaptosuccinic acid scan images using DCT and discussed modality-dependent statistical characteristics that differ from conventional radiological images. Their results suggest that DCT-based coding can be adapted to nuclear medicine data under near-lossless fidelity requirements.

\textbf{Wavelet-based Methods}.
Wavelet-based codecs follow the general framework in Fig.~\ref{fig:2dtransform} and use multiresolution wavelet decomposition to produce subbands across scales and orientations. Compared with block-based transforms, wavelet representations can reduce blocking artifacts and support embedded and scalable coding through coefficient organization and bit-plane based entropy coding.

The theoretical foundation of wavelet-based representation is commonly attributed to Mallat's multiresolution analysis, which formalized wavelet representations and showed that signals can be decomposed into a coarse approximation and detail components across scales while preserving perfect reconstruction~\cite{mallat2002theory}.

DeVore, Jawerth, and Lucier developed an analysis framework for wavelet-transform coding, relating approximation error decay to function smoothness characterized in Besov-type spaces and clarifying how quantization of wavelet coefficients affects reconstruction error under these models~\cite{devore2002image}.

Lewis and Knowles presented an early practical implementation of 2D wavelet-based image compression by combining orthogonal wavelet decomposition with hierarchical coefficient coding and perceptual modeling motivated by the human visual system. They reported competitive performance against contemporaneous transform and vector-quantization baselines under their settings~\cite{lewis1992image}.

A major step toward embedded wavelet coding was Shapiro's Embedded Zerotree Wavelet algorithm. It exploited cross-scale self-similarity via zerotree significance relationships to generate an embedded bitstream that supports progressive reconstruction and rate control, and it was commonly combined with successive-approximation quantization and adaptive arithmetic coding~\cite{shapiro2002embeddedsaid1996new}.

Building on EZW, Said and Pearlman proposed SPIHT, which uses a more efficient set-partitioning strategy to code wavelet coefficients and reduces the need for explicit zerotree symbol transmission. They reported improved rate-distortion efficiency with lower computational and memory overhead while preserving the embedded property~\cite{said1996}.

From a standardization perspective, JPEG 2000 is a representative wavelet-based coding system for still images, supporting both lossless and lossy compression, progressive transmission, and region-of-interest capabilities~\cite{skodras2001jpeg}. In medical imaging workflows, DICOM specifies JPEG 2000 transfer syntaxes. JPEG 2000 also supports reversible 5/3 and irreversible 9/7 wavelet transformations, making wavelet-domain compression directly relevant to clinical storage and transmission pipelines~\cite{dicom_standard}.

Recent studies continue to explore wavelet-based designs tailored to medical constraints. Ammah and Owusu proposed a hybrid DWT--VQ approach and discussed maintaining medically acceptable perceptual quality under their evaluated settings~\cite{ammah2019robust}. Farghaly and Ismail investigated DWT-based compression optimized for FPGA implementation to support real-time processing requirements~\cite{farghaly2020floating}. Viswanathan and Palanisamy performed an empirical study on wavelet selection for near-lossless medical image compression and evaluated different wavelets under a SPIHT-based pipeline~\cite{viswanathan2024empirical}.

\textbf{Other tranform-based Methods}.
Beyond DCT and wavelet families, medical image compression has also explored statistical transforms, geometric multiscale representations, projection-based transforms, and newly designed transforms tailored to specific medical constraints such as region-of-interest fidelity, edge preservation, and modality-dependent noise characteristics.

Statistical transform approaches leverage data-driven decorrelation, with principal component analysis being a representative example that can provide strong energy compaction under a second-order statistical model. Lim et al.~\cite{ting2015novel} introduced a PCA-based method for arbitrary-shape region-of-interest compression in medical images. Their design aimed to preserve diagnostically important regions while allowing higher compression in less critical areas, illustrating how ROI prioritization can be integrated into a transform-based pipeline.

Building on statistical transforms, Reddy et al.~\cite{reddy2018new} developed a hybrid PCA--SPIHT approach that combines PCA for decorrelation with SPIHT for embedded progressive coding. They reported improved reconstruction quality and higher PSNR under their experimental settings compared with several baselines, including DCT-based and PCA-based variants, and discussed its potential utility for storage and telemedicine transmission.

To better represent anisotropic and geometric structures such as edges and contours that are common in medical imagery, geometric multiscale analysis methods have also been investigated. Anandan et al.~\cite{anandan2016medical} proposed medical image compression using a wrapping-based fast discrete curvelet transform with arithmetic coding. Their study highlighted that curvelet representations can capture directional edge information more effectively than separable transforms under their evaluated settings, which is relevant for images with complex boundaries.

Contourlet transforms have likewise been explored for medical modalities with strong directional content and noise. Hashemi-Berenjabad and Mahloojifar~\cite{hashemi2013new} developed a contourlet-based compression and speckle reduction method for ultrasound images. They reported improved preservation of image details and edges under high speckle levels compared with wavelet-based baselines, while achieving effective compression performance and providing practical guidance on contourlet parameter selection.

Projection-based approaches exploit directional information through transforms defined on projections. Juliet et al.~\cite{juliet2015projection} presented a Radon transform-based compression system designed for telemedicine applications. Their method leveraged the dimension-reducing property of Radon projections to convert 2D processing into a set of 1D operations and reported competitive performance compared with several transform-based alternatives under their evaluation settings.

Recent work has also investigated transforms designed specifically for lossless medical image compression rather than adapting existing transforms. Rojas-Hernández et al.~\cite{rojas2022lossless} introduced a difference transform for lossless compression and reported gains over JPEG 2000 under their evaluated datasets and settings, while aiming for a relatively simple implementation. This line of work suggests that application-specific transform design can still be beneficial for particular medical imaging scenarios.

Finally, practical adoption is strongly influenced by standards and workflow constraints. Liu et al.~\cite{liu2017current} reviewed the role of image compression standards in medical imaging and discussed how different approaches align with clinical requirements such as interoperability, archival storage, and transmission, providing practical considerations for selecting codecs in real-world deployments.

\begin{figure}[htbp]
	\centering
	\includegraphics [width=\linewidth]{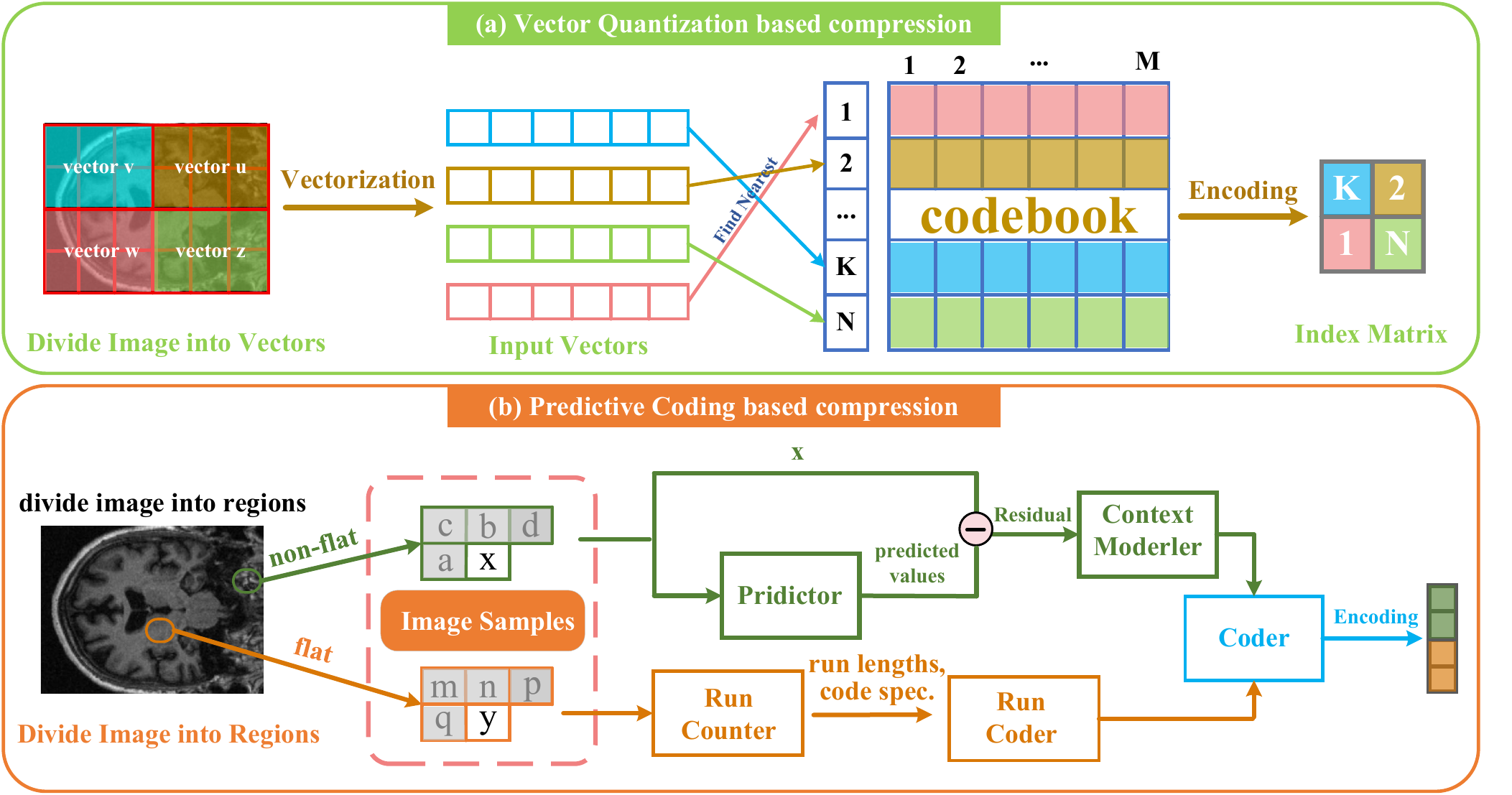}\\
	\caption{Architectural Comparison of Vector Quantization and Predictive Coding Methods for 2D Medical Image.}
	\label{fig:2D_tradi_VQ_PC}
    \vspace{-0.3cm}
\end{figure}

\subsubsection{Predictive Coding Methods}
While transform-based methods have achieved strong performance in medical image compression, 2D medical image compression also includes two other fundamental families that follow different principles: predictive coding and vector quantization. Fig.~\ref{fig:2D_tradi_VQ_PC} summarizes their architectural differences. \rvnew{Predictive-coding-based compression exploits spatial redundancy by predicting each pixel from previously processed neighbors and encoding only the prediction residuals using context modeling and entropy coding. This family is widely used in lossless and near-lossless settings, where preserving subtle diagnostic details is critical.}

Classical context-based adaptive algorithms in the 1990s established the foundation for predictive coding. Weinberger et al.~\cite{weinberger1996loco} introduced LOCO-I, which combines a simple predictor with context modeling and entropy coding and achieves strong lossless compression with low complexity. The method was designed to work efficiently with adaptive Golomb-type codes and also includes mechanisms for handling low-entropy regions.

Building on these principles, Weinberger et al.~\cite{weinberger2000loco} standardized the approach in JPEG-LS, which supports lossless and near-lossless compression of continuous-tone images and is relevant to medical workflows that require interoperability. In the same period, Wu and Memon~\cite{wu1996calic} developed CALIC, a context-based adaptive lossless codec that has been widely used as a high-performance baseline for subsequent predictive coding research.

Later work explored how to adapt predictive coding pipelines to medical image characteristics and constraints. Shirsat and Bairagi~\cite{shirsat2013lossless} combined integer wavelet transforms with prediction to improve lossless compression performance for medical images, aiming to better capture structural characteristics while preserving diagnostic quality.

Puthooran et al.~\cite{puthooran2013lossless} proposed a dual-level differential pulse code modulation approach for lossless medical image compression. The method applies a linear DPCM stage followed by a context-adaptive switching neural network predictor, which selects among multiple predictors based on local texture context. They reported improved compression performance on MRI images compared with common baselines such as CALIC and LOCO-I under their evaluation settings.

Predictive coding has also been extended to medical image sequences by incorporating temporal redundancy. Miaou et al.~\cite{miaou2009lossless} combined JPEG-LS with interframe coding to exploit both spatial and temporal correlations for improved compression of medical image sequences.

Because medical images often contain regions with different diagnostic importance, ROI mechanisms have been integrated with prediction-based coding. Zuo et al.~\cite{zuo2015improved} incorporated lossless ROI functionality to preserve critical regions with perfect fidelity while applying different strategies to non-ROI areas. Eben Sophia and Anitha~\cite{eben2016region} further proposed a region-based prediction scheme with quality measurements tailored to medical images, motivated by the observation that diagnostically relevant ROIs may occupy only a small fraction of the image.

Gradient-adaptive prediction and residual processing have also been studied to improve coding efficiency. Baware et al.~\cite{baware2016medical} introduced a predictor adaptive to four directional gradients, selecting the direction with the least gradient to improve prediction accuracy. Before entropy coding, residuals were grouped using a maxplane-based strategy to further enhance coding efficiency, and the authors reported improved compression ratios compared with CALIC and basic DPCM variants under their settings.

Optimization strategies were later introduced to refine prediction and reduce large residual errors. Anitha et al.~\cite{anitha2018optimized} proposed an optimized predictive coding method that generates a binary mask using optimized thresholds and performs prediction on masked coefficients to reduce high-error contributions. They reported up to 45\% improvement in compression ratio compared with a conventional prediction process under their evaluation settings.

Data-driven techniques have also been integrated into predictive pipelines. Xin et al.~\cite{xin2021lossless} employed data mining to build image codebooks by extracting basic components from multi-component medical images. Their method decomposes multi-component images into single-component images, applies a reversible component transform, and then performs predictive coding. A subsequent mapping step separates images into shape and detail layers, and the authors reported improved performance compared with PNG and JPEG2000 under their tested settings.


\subsubsection{Vector Quantization Methods} 
\rvnew{Vector-quantization-based compression, shown in Fig.~\ref{fig:2D_tradi_VQ_PC}a, partitions an image into non-overlapping blocks represented as vectors and maps each vector to its closest entry in a trained codebook via index assignment. Compression is achieved by replacing the original vectors with compact indices, with the compression ratio determined by the codebook size and vector dimension. A typical VQ pipeline includes vector formation, codebook training, index assignment, and entropy coding of indices, and many medical designs further incorporate ROI handling or adaptive block partitioning to protect diagnostically important structures.}

VQ is a widely studied lossy technique for medical image compression, particularly when images contain repetitive patterns that can be represented by shared codebook entries. Classical codebook design is often associated with the LBG algorithm~\cite{linde2003algorithm}, and later work has focused on addressing limitations that are important in medical imaging, such as edge preservation and ROI fidelity.

Early work by Cazuguel et al.~\cite{cziho1998medical} introduced ROI-oriented vector quantization for medical images, highlighting the need to prioritize diagnostically relevant regions during compression. This line of work motivated adaptive strategies that allocate more bits to ROI areas while allowing stronger quantization elsewhere.

Jiang et al.~\cite{jiang2012medical} proposed a wavelet-domain VQ framework with adaptive block sizes to mitigate edge degradation in conventional VQ. Their method used local fractal dimension to estimate local complexity, partitioned high-complexity regions into smaller blocks and low-complexity regions into larger blocks, and adopted a modified K-means clustering strategy for codebook training. They reported improved reconstruction quality under their evaluation settings compared with fixed-block VQ variants.

Related wavelet-based adaptive VQ designs have also been explored. Mitra et al.~\cite{mitra1998wavelet} studied quantization of wavelet-decomposed subimages using clustering strategies and reported high compression ratios under their tested settings, indicating the potential of combining multiresolution representations with VQ-style quantization.

Optimization-based codebook design has been investigated to further improve VQ performance for medical imaging. Kumar et al.~\cite{kumar2018compression} applied simulated annealing for codebook optimization in contextual vector quantization for CT image compression. Their pipeline separated ROI and background using region-growing segmentation and encoded the two regions with different bit allocations. They reported improvements in PSNR, MSE, and compression ratio compared with several baselines under their experimental settings.

Wavelet--VQ hybrids have also been motivated by the energy compaction of wavelets and the quantization efficiency of VQ. Gurjar and Korde~\cite{gurjar2014medical} evaluated combinations of several wavelet families with vector quantization for telemedicine-oriented scenarios and discussed how wavelet choice affects compression behavior under their settings.

\color{revblue}
\subsubsection{Run-Length Encoding Methods}
\rvnew{
Run-Length Encoding (RLE) is a simple lossless compression technique that represents consecutive identical values using the value and its repetition count~\cite{salomon2002data}. As a lossless method, RLE introduces no quantization error. However, it does not explicitly remove spectral redundancy, and its effectiveness depends on the presence of long runs, which are often disrupted by noise and rich anatomical textures in continuous-tone modalities such as CT and MRI.}

\rvnew{
RLE remains relevant in medical imaging mainly due to standardization and workflow compatibility. To partially mitigate the run-breaking effect of texture and noise, hybrid variants have been explored, for example locally optimal RLE for CT images~\cite{rhodes2007locally}. In particular, DICOM defines an RLE lossless transfer syntax for pixel data encapsulation~\cite{dicom2021part5,mildenberger2002introduction}. Beyond raw images, RLE is especially effective for sparse or piecewise-constant derived data such as segmentation masks, overlays, and annotations, where large homogeneous regions naturally yield long runs~\cite{wasserthal2023totalsegmentator,lin2014microsoft}. Consequently, in modern pipelines, RLE is typically used as an auxiliary coder for masks and metadata or as a component in hybrid schemes rather than a standalone compressor for raw anatomical scans.}

\subsection{Learning based Medical Images Compression Methods}

\color{revblue}
\subsubsection{Evolution of Learning-Based Medical Image Compression}

Building on the taxonomy discussed in previous sections, learning-based medical image compression has progressed through multiple generations of neural architectures rather than a single paradigm. Existing approaches can be organized into five families: convolutional autoencoder based codecs, recurrent and ConvLSTM sequence models, attention and Transformer architectures, generative compression frameworks, and implicit neural representation with hybrid schemes~\cite{ma2019image,bourai2024deep,molaei2023implicit}. These families overlap in time, and each emerged to address limitations of earlier designs and to better exploit spatial redundancy, volumetric continuity, and modality-dependent statistics in medical data.

Most learned codecs share a common pipeline. An encoder maps the input to a compact latent representation, which is quantized for bitstream generation. A probabilistic entropy model estimates the likelihood of quantized latents, enabling arithmetic coding, and a decoder reconstructs the image or volume from the decoded latents. Different families mainly differ in how the encoder and decoder represent anatomy and how the entropy model captures context across space, slices, and time, often under constraints such as diagnostic fidelity, anisotropic sampling, and clinical workflow compatibility.

\begin{figure*}[!t]
	\centering
	\includegraphics [width=0.8\textwidth]{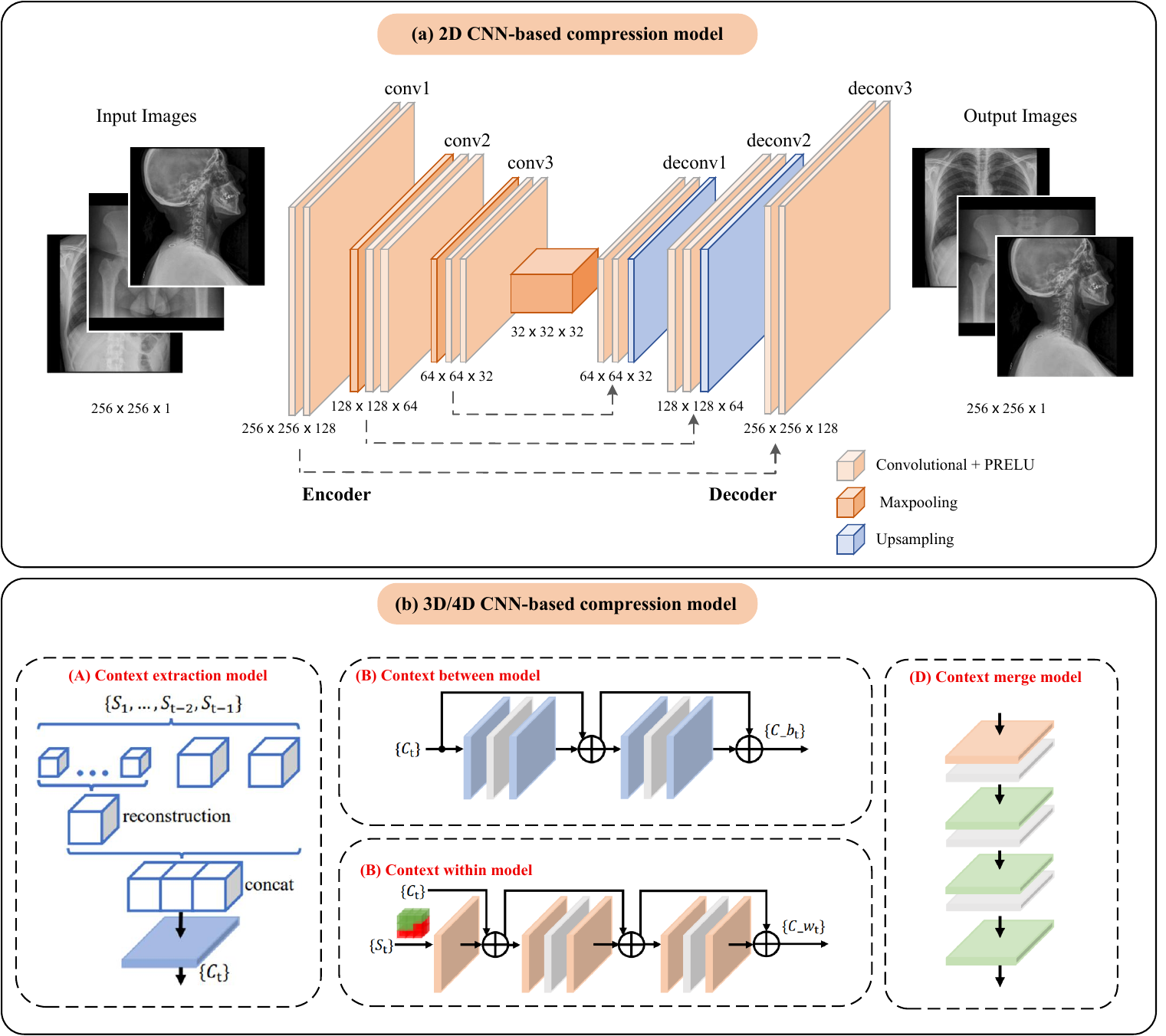}\\
	\caption{Comparison of CNN-based compression architectures for 2D and 3D medical images. (a) 2D CNN-based compression methods \cite{fettah2024convolutional} employ conventional encoder-decoder architectures, where input images undergo compression through encoder networks and subsequent reconstruction via decoder networks. (b) 3D CNN-based compression methods \cite{xue2022aiwave} exploit inter- and intra-subband correlations through hierarchical context modeling to reduce coding uncertainty and achieve improved compression efficiency for volumetric medical data.}
	\label{fig:2DVS3D_CNNbased_1}
    \vspace{-0.5cm}
\end{figure*}

The first generation is dominated by convolutional autoencoder based codecs and has remained a backbone of many medical designs~\cite{liu2022medical,fettah2024convolutional}. These methods use convolutional encoders and symmetric decoders to compress 2D images or individual slices. Their locality and weight sharing capture intra-slice spatial redundancy in radiography, ultrasound, and histopathology. Fig.~\ref{fig:2DVS3D_CNNbased_1} illustrates typical CNN-based pipelines for 2D and 3D medical data. While CAE-based methods often achieve strong rate--distortion performance in 2D settings, they primarily model local context and can be limited in capturing long-range anatomical dependencies and volumetric consistency.

To better exploit inter-slice and temporal redundancies, recurrent and ConvLSTM based models were introduced and have been extensively explored for 3D and 4D medical data~\cite{nagoor2021medzip,chen2023streaming,nagoor2022sampling,sridhar2022optimal}. These approaches treat CT or MRI slices and dynamic frames as sequences and propagate state along the slice or time dimension, strengthening context modeling beyond purely 2D codecs. However, sequential processing introduces latency and limits parallelism, and the effective context across long sequences can remain constrained.

More recently, attention and Transformer based codecs have become a central trend in both 2D and volumetric medical image compression~\cite{ma2019image,liu2022learning,bourai2024deep}. Self-attention relaxes the locality of convolution and the strict ordering of RNNs, enabling direct modeling of long-range correlations within and across slices. Transformer codecs often report improved rate--distortion performance compared with CNN and RNN baselines, but with higher computational and memory costs, motivating windowed, hierarchical, and low-rank attention variants tailored to large medical volumes~\cite{bourai2024deep}.

In parallel, generative model based compression explores stronger learned priors to reconstruct high-frequency details from very compact representations. GAN-based codecs for natural images demonstrate visually pleasing reconstructions at extremely low bit rates~\cite{agustsson2019generative,iwai2021fidelity}. In the medical domain, early VAE-based codecs~\cite{liu2022medical} and several deep-learning-assisted compression frameworks and studies~\cite{aziz2020comprehensive,fettah2024deep,bourai2024deep} adopt probabilistic formulations that treat reconstruction as sampling or generation from low-dimensional latents. These schemes can be effective at low bit rates, but they raise concerns about hallucinated structures, loss of subtle diagnostic cues, and the difficulty of aligning perceptual metrics with clinical endpoints, which limits deployment in safety-critical workflows~\cite{bourai2024deep}.

A fifth family extends beyond grid-based representations through implicit neural representation and hybrid designs~\cite{sitzmann2020implicit,mildenhall2021nerf,dupont2021coin,molaei2023implicit}. INR methods encode an image or volume as a continuous coordinate-to-intensity function parameterized by a compact network, supporting flexible resampling and potentially compact storage via network parameters~\cite{ma2024semantic,yang2023sci,dai2024implicit,gao2023sinco}. Hybrid schemes combine traditional components such as transforms, predictive coding, or standard codecs with learned modules for feature extraction, entropy modeling, or ROI refinement, seeking to inherit robustness and standard compatibility while leveraging adaptivity of deep networks~\cite{xue2022aiwave,zheng2024hybrid,li2025towards,chen2022exploiting}.

Overall, learning-based medical image compression has progressed from local convolutional models to sequence-aware architectures, then to Transformer-based codecs, and further to generative and INR-based paradigms that treat compression as conditional synthesis or continuous function approximation~\cite{ma2019image,liu2022learning,bourai2024deep,molaei2023implicit}. Key directions include scaling attention-based models to large 3D and 4D studies, improving clinical reliability of generative compression, and unifying INR and hybrid codecs with pretraining strategies while maintaining compatibility with existing medical imaging infrastructures~\cite{bourai2024deep,molaei2023implicit}.
\color{black}

\subsubsection{Neural Network-based Codec Methods}
\rvnew{
Neural network-based codec methods have become an important direction in medical image compression by learning compact representations directly from data. Unlike traditional codecs that rely on handcrafted transforms and fixed statistical assumptions, these approaches train encoder--decoder networks end to end to optimize a rate--distortion objective. A typical learned codec maps an image or slice to a latent representation, applies quantization to enable bitstream generation, models the probability of quantized latents for arithmetic coding, and reconstructs the image via a neural decoder. Domain constraints such as diagnostic fidelity, structural preservation, and ROI prioritization can be incorporated through architecture design and loss formulations.}

Fig.~\ref{fig:2DVS3D_CNNbased_1} highlights key challenges when extending 2D neural codecs to volumetric data. While 2D designs often follow a standard encoder--decoder structure, 3D methods must additionally model correlations along the depth dimension, which motivates explicit context extraction and modeling modules. Representative volumetric designs therefore aim to capture both intra-slice and inter-slice dependencies to improve coding efficiency for 3D anatomy.

Liu et al.~\cite{liu2022medical} proposed a VAE-based medical image compression framework that integrates residual modules and optimizes the rate--distortion trade-off within a probabilistic formulation. Their approach models latent distributions of medical image features and is trained to preserve diagnostically relevant information under a learned prior. Fettah et al.~\cite{fettah2024convolutional} introduced the Medical X-ray Imaging Dataset (MXID) and provided a systematic evaluation of compression techniques ranging from traditional baselines to deep learning models, including CNNs, DCAEs, AEs, and VAEs. Their study reported that CAE-based codecs can preserve anatomical structures with competitive rate--distortion performance under their evaluated settings, offering useful benchmarks for medical image compression.

\color{revblue}
Recently, Vikraman et al.~\cite{vikraman2024segmentation} proposed a segmentation-driven compression approach for brain MRI. Their method separates the image into ROI and non-ROI regions, compresses the ROI using an optimized CNN, and encodes the non-ROI region using an RNN. The authors reported a PSNR of 45.502 under their experimental settings and discussed improved preservation of diagnostically relevant regions for high-resolution MRI data.
\color{black}

\subsubsection{Implicit Neural Representation-based Methods}
\rvnew{
Implicit Neural Representation (INR) methods compress images by representing them as continuous coordinate-to-intensity functions parameterized by a neural network rather than as discrete pixel arrays. A typical INR compression pipeline samples coordinates, applies positional encoding, fits a compact multilayer perceptron to map encoded coordinates to pixel or voxel values, and stores a compact description of the fitted model, such as network weights or lightweight modulation parameters. At decoding time, the image is reconstructed by querying the network at desired coordinates, which naturally supports flexible resampling and resolution changes.}

COIN~\cite{dupont2021coin} and COIN++~\cite{dupont2022coin++} are representative early works in INR-based image compression. COIN fits a multilayer perceptron to map pixel coordinates to RGB values and stores network parameters rather than raw pixels, reporting improved performance over JPEG under low-bit-rate settings. COIN++ extends this idea with meta-learning and modulation-based compression by using a shared base network and storing compact modulations for each signal, aiming to reduce per-image encoding cost and improve encoding speed. This formulation also illustrates how coordinate-based models can be applied across different data types under a unified representation.

Subsequent work has improved INR compression through better training strategies and architectural choices. Str{\"u}mpler et al.~\cite{strumpler2022implicit} proposed a pipeline with meta-learned initialization and sinusoidal representation networks with positional encoding, and reported gains over JPEG2000 under their evaluated settings. Additional architectural and optimization advances have been explored to improve rate--distortion performance~\cite{sanjeet2024breaking}. INR-based compression has also been investigated in scientific and biomedical imaging settings, indicating its potential beyond natural image benchmarks~\cite{rezasoltani2023hyperspectral,chen2021nerv}.

\begin{figure*}[t]
	\centering
	\includegraphics [width=0.85\textwidth]{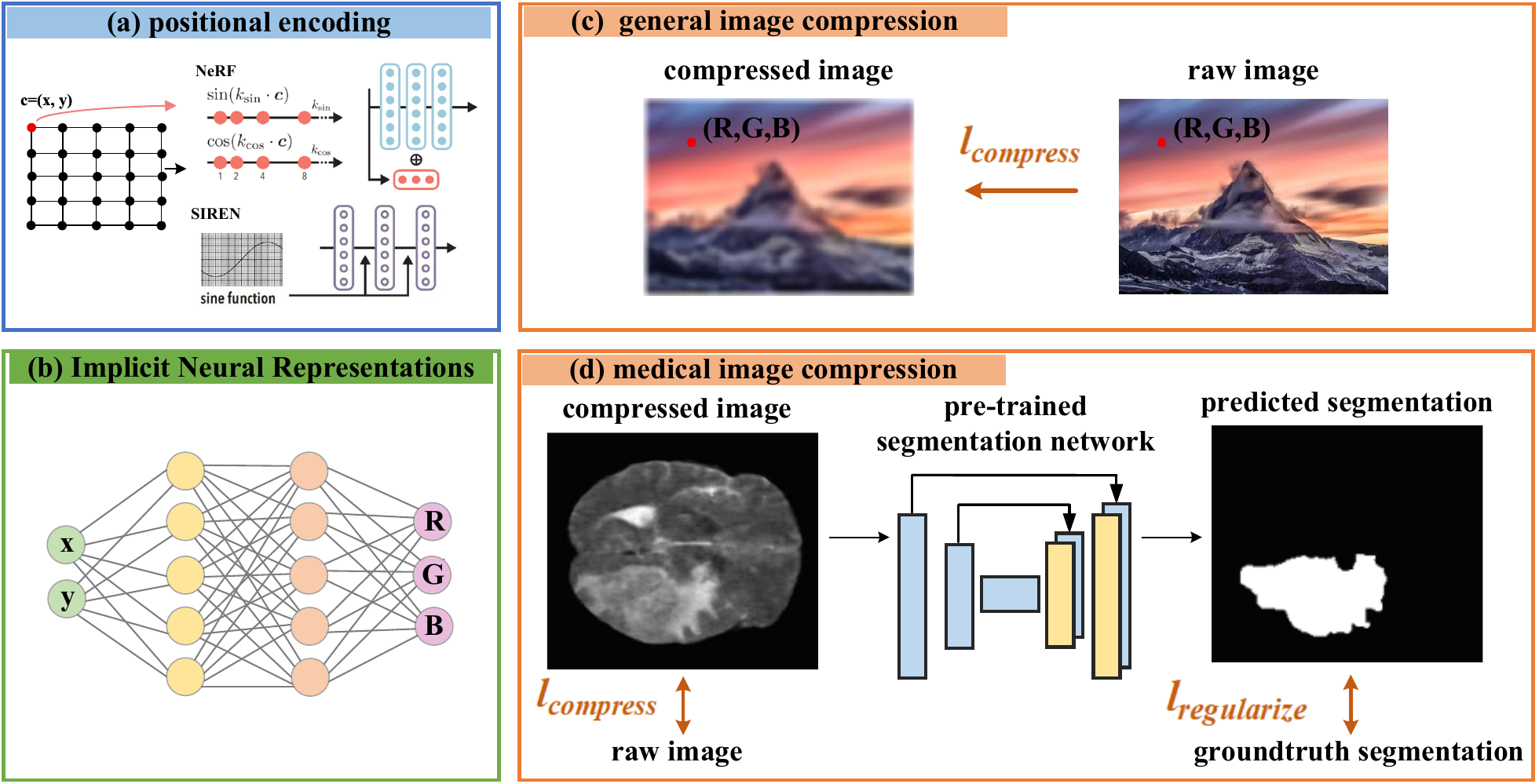}\\
	\caption{\textbf{Comparison of 2D INR methods for general image and medical image compression \cite{gao2023sinco}.} (a) Use position encoding strategies such as NeRF and SIREN to encode the coordinate of pixels. (b) Multi-layer perceptron that learns coordinate-to-pixel mappings from encoded positions to RGB values. (c) General image compression employs standard reconstruction loss between compressed and raw images. (d) Medical image compression incorporates structural regularization through a pre-trained segmentation network, combining pixel-wise loss with segmentation consistency loss to preserve anatomical structures.}
        \label{fig:2DINRgeneralVSmedical}
        \vspace{-0.5cm}
\end{figure*}

Fig.~\ref{fig:2DINRgeneralVSmedical} illustrates representative differences between general-purpose and medical INR compression designs. INR models typically rely on positional encoding strategies, such as sinusoidal encodings and sine-based activations, to better capture high-frequency details, and use an MLP to map encoded coordinates to pixel values. In medical imaging, a key additional requirement is structural fidelity, where preserving anatomically relevant boundaries and regions is often more important than optimizing generic perceptual similarity.

Accordingly, medical INR compression may incorporate structural regularization in the loss design. Beyond pixel-wise reconstruction loss, a pre-trained segmentation network can be used to enforce consistency of anatomical structures between the original and reconstructed images by adding a segmentation-based regularization term.

Building on this idea, Gao et al.~\cite{gao2023sinco} proposed SINCO, which introduces a structural regularizer for INR-based compression using segmentation guidance. Their method couples an INR model with a segmentation network and optimizes a Dice-based consistency objective between segmentation maps of the original and reconstructed images, aiming to better preserve anatomically important structures under compression.

\newlist{tabitemize}{itemize}{1}
\setlist[tabitemize]{
    label=\textbullet,
    leftmargin=*,        
    nosep,               
    topsep=3pt,          
    partopsep=0pt,
    parsep=0pt,
    itemsep=3pt,         
    after=\vspace{2pt}   
}

\definecolor{headerbg}{RGB}{0, 50, 100}         
\definecolor{col_sidebar}{RGB}{235, 242, 250}   
\definecolor{row_odd}{RGB}{255, 255, 255}       
\definecolor{row_even}{RGB}{248, 250, 251}      

\newcolumntype{M}[1]{>{\RaggedRight\arraybackslash}m{#1}}
\newcommand{\head}[1]{\multicolumn{1}{c}{\textbf{\textcolor{white}{#1}}}}

\begin{table*}[t]
\centering
\caption{Qualitative Comparison of Representative 2D Medical Image Compression Methods}
\label{tab:2d_methods}

\renewcommand{\arraystretch}{1.3} 
\setlength{\tabcolsep}{5pt}        
\setlength{\aboverulesep}{0pt}
\setlength{\belowrulesep}{0pt}
\arrayrulecolor{headerbg}

\begin{tabular}{
    >{\columncolor{col_sidebar}\bfseries}M{2.8cm} 
    M{3.2cm}   
    M{3.4cm}  
    M{3.4cm}   
    M{3.4cm}   
}

\toprule

\rowcolor{headerbg} 
\head{Category} & 
\head{Representative Works} & 
\head{Key Characteristics} & 
\head{Strengths} & 
\head{Limitations} \\

\rowcolor{row_even}
Transform-based &
DWT--DCT hybrid~\cite{singh2007dwt}; \newline
DCT subband + SPIHT~\cite{chen2007medical}; \newline
Curvelet-based~\cite{anandan2016medical} &
\begin{tabitemize}
    \item Block/subband coding
    \item Entropy coding optimization
    \item ROI-oriented support
\end{tabitemize} &
\begin{tabitemize}
    \item Mature technology
    \item Good R-D performance
    \item Near-lossless control
    \item Fast decoding
\end{tabitemize} &
\begin{tabitemize}
    \item Blocking/ringing artifacts
    \item Fixed bases miss complex anatomy
    \item Sensitive to parameters
\end{tabitemize} \\

\rowcolor{row_odd}
Predictive Coding &
Lossless schemes~\cite{shirsat2013lossless}; \newline
ROI-adaptive pred.~\cite{zuo2015improved} &
\begin{tabitemize}
    \item Pixel/block prediction 
    \item Edge-directed modeling
    \item Often supports lossless ROIs
\end{tabitemize} &
\begin{tabitemize}
    \item Low complexity \& memory
    \item Suitable for low-resource devices
    \item Reliable lossless ROI preservation
\end{tabitemize} &
\begin{tabitemize}
    \item Limited capacity for textures
    \item Lower efficiency than DL methods
    \item Performance depends on edge detection
\end{tabitemize} \\

\rowcolor{row_even}
Vector Quantization &
ROI-VQ~\cite{cziho1998medical}; \newline
Wavelet-VQ~\cite{mitra1998wavelet}; \newline
Variable block VQ~\cite{jiang2012medical} &
\begin{tabitemize}
    \item Codebook-based quantization
    \item Trained on specific modalities
    \item Block or subband domain
\end{tabitemize} &
\begin{tabitemize}
    \item Simple decoding structure
    \item Modality-specific tuning
    \item Wavelet-VQ reduces blocking
\end{tabitemize} &
\begin{tabitemize}
    \item Training is costly \& data-dependent
    \item Blocking artifacts at high CR
    \item Poor generalization
\end{tabitemize} \\

\rowcolor{row_odd}
Run-Length Encoding &
RLE for CT~\cite{rhodes2007locally}; \newline
JPEG-LS sequences~\cite{miaou2009lossless}; \newline
DICOM syntaxes~\cite{mildenberger2002introduction} &
\begin{tabitemize}
    \item RLE/Difference coding
    \item Embedded in DICOM syntaxes
    \item Often combined with JPEG-LS
\end{tabitemize} &
\begin{tabitemize}
    \item Extremely fast
    \item Very low complexity
    \item Fully DICOM compatible
    \item Legal archiving safe
\end{tabitemize} &
\begin{tabitemize}
    \item Low compression ratios
    \item No spatial correlation modeling
    \item Weak for storage constraints
\end{tabitemize} \\

\rowcolor{row_even}
Neural Network-based &
VAE-based~\cite{liu2022medical}; \newline
CNN+RNN~\cite{vikraman2024segmentation}; \newline
DWT+CNN~\cite{paul2022health} &
\begin{tabitemize}
    \item Learn latent representations
    \item End-to-end optimization
    \item Uses Conv-AEs or VAEs
\end{tabitemize} &
\begin{tabitemize}
    \item Outperforms hand-crafted codecs
    \item Adapts to institution statistics
    \item Task-specific optimization
\end{tabitemize} &
\begin{tabitemize}
    \item Compute-intensive training
    \item Risk of overfitting or hallucination
    \item Lack of standardization
\end{tabitemize} \\

\rowcolor{row_odd}
INR-based &
SinCo (Brain MRI)~\cite{gao2023sinco} &
\begin{tabitemize}
    \item Continuous functions 
    \item Coordinate-to-intensity mapping
    \item Uses biomedical priors
\end{tabitemize} &
\begin{tabitemize}
    \item Resolution-independent
    \item Continuous zoom support
    \item Sub-pixel accuracy
\end{tabitemize} &
\begin{tabitemize}
    \item Slow training per instance
    \item Hard to scale to archives
    \item Clinical validation pending
\end{tabitemize} \\

\rowcolor{row_even}
Hybrid Methods &
RNN+GenPSOWVQ~\cite{sridhar2022optimal} &
\begin{tabitemize}
    \item Classical transforms + Learned modules
    \item Joint feature/entropy modeling
\end{tabitemize} &
\begin{tabitemize}
    \item Robustness of classical codecs
    \item Adaptivity of deep learning
    \item DICOM-compatible designs
\end{tabitemize} &
\begin{tabitemize}
    \item Complex system design
    \item Coordinating components is hard
    \item Regulatory difficulty
\end{tabitemize} \\

\bottomrule

\end{tabular}
\end{table*}

\definecolor{headerbg}{RGB}{0, 50, 100}         
\definecolor{col_sidebar}{RGB}{235, 242, 250}   
\definecolor{row_odd}{RGB}{255, 255, 255}       
\definecolor{row_even}{RGB}{248, 250, 251}      

\begin{table*}[t]
\centering
\caption{Quantitative comparison of representative 2D medical image compression methods (values reported from original papers).
The ``Dataset / Experimental Setting'' column summarizes the data source (DS), modality, and bit-depth (BD), together with key characteristics explicitly stated in the original works.
Experimental conditions (datasets, bit-depths, evaluation protocols, and compression modes) vary across references; therefore, the comparison is indicative rather than a strictly controlled benchmark.}
\label{tab:quant_2D}
\footnotesize

\renewcommand{\arraystretch}{1.15}
\setlength{\tabcolsep}{5pt}
\arrayrulecolor{headerbg}

\begin{tabular}{
    >{\columncolor{col_sidebar}\bfseries}c
    c
    l
    l
    >{\raggedright\arraybackslash}p{3.8cm}
    c
    c
    c
}

\toprule

\rowcolor{headerbg}
\head{Ref.} &
\head{Year} &
\head{Type} &
\head{Method} &
\head{Dataset / Experimental Setting} &
\head{Bpp} &
\head{PSNR} &
\head{SSIM} \\

\rowcolor{row_even}
\cite{rhodes2007locally} & 1985 & Lossless & Locally optimal RLE
& \makecell[l]{DS=CT slices (2D)\\ BD=12b, CR$\approx$0.525}
& -- & -- & -- \\

\rowcolor{row_odd}
\cite{mitra1998wavelet} & 1998 & Lossy & Wavelet-based VQ 
& \makecell[l]{DS=Cervical spine radiographs\\ BD=8b, 1024$\times$1024}
& 0.18 & 41.98 & -- \\

\rowcolor{row_even}
\cite{chen2007medical} & 2007 & Lossy & DCT-CSPIHT
& \makecell[l]{DS=Angiogram (2D)\\ BD=8b, 512$\times$512}
& 0.10 & 45.20 & -- \\

\rowcolor{row_odd}
\cite{jiang2012medical} & 2012 & Lossy & Wavelet-domain VQ 
& \makecell[l]{DS=Liver \& brain CT slices\\ BD=12--16b, 512$\times$512; CR=25}
& -- & 37.50 & -- \\

\rowcolor{row_even}
\cite{shirsat2013lossless} & 2013 & Lossless & IWT + Predictive coding
& \makecell[l]{DS=X-ray / CT / MRI (2D)\\ BD=NS, CR $\approx$ 1.3--1.8}
& -- & -- & -- \\

\rowcolor{row_odd}
\cite{zuo2015improved} & 2015 & Lossy & IMIC-ROI
& \makecell[l]{DS=CT / MRI / angiogram (2D)\\ BD=NS, ROI-lossless hybrid}
& -- & -- & 0.9988 \\

\rowcolor{row_even}
\cite{anandan2016medical} & 2016 & Lossy & FDCVT + arithmetic coding
& \makecell[l]{DS=CT brain image (2D)\\ BD=NS; CR$\approx$55.6}
& -- & 35.75 & -- \\

\rowcolor{row_odd}
\cite{paul2022health} & 2022 & Lossy & MIC-DWT-CNN
& \makecell[l]{DS=MedPix \\ BD=NS, ROI-lossless hybrid}
& -- & 35.11 & -- \\

\rowcolor{row_even}
\cite{sridhar2022optimal} & 2022 & Lossy & RNN + GenPSOWVQ
& \makecell[l]{DS=CT / X-ray / mammogram \\ BD=NS, fixed CR=25\%}
& -- & 60.64 & 0.86 \\

\rowcolor{row_odd}
\cite{gao2023sinco} & 2023 & Lossy & INR (SINCO)
& \makecell[l]{DS=Brain MRI (Decathlon)\\ BD=NS, fixed bpp=1.2}
& 1.20 & 35.69 & 0.952 \\

\rowcolor{row_even}
\cite{vikraman2024segmentation} & 2024 & Lossy & CNN + RNN
& \makecell[l]{DS=Brain MRI (Kaggle LGG)\\ BD=NS, segmentation-guided}
& -- & 45.50 & -- \\

\bottomrule
\end{tabular}
\end{table*}

\subsubsection{Hybrid Methods}
\rvnew{
Hybrid methods in medical image compression combine traditional signal processing components with learned modules to exploit complementary strengths~\cite{bourai2024deep}. In a typical hybrid pipeline, classical tools such as transforms, prediction, or standard codec structures provide a stable representation and efficient baseline coding, while neural networks are introduced to improve specific stages such as feature extraction, context and entropy modeling, ROI-aware bit allocation, or reconstruction enhancement~\cite{paul2022health}. Many studies report that such designs can achieve better rate--distortion trade-offs than purely traditional or purely learning-based baselines under their evaluated settings~\cite{li2025towards}.}

Paul and Chandran~\cite{paul2022health} proposed a healthcare image compression scheme that combines discrete wavelet transform with convolutional neural networks, illustrating a wavelet--CNN hybrid design. Their study reported improved compression performance under their settings while maintaining compatibility with a wavelet-domain representation.

Li et al.~\cite{li2025towards} presented a scalable hybrid framework that combines DWT with CNN architectures. Their SDWTCNN design uses a transform-based representation to support ROI-aware coding and applies learned modules to enhance compression for other regions, with additional components such as SVD-based feature extraction for ROI processing. The authors reported improved compression performance under their evaluation settings and discussed practical considerations on complexity and scalability.

Another category of hybrid methods combines learned feature modeling with traditional optimization and quantization strategies. Sridhar et al.~\cite{sridhar2022optimal} proposed a two-stage approach that first uses an RNN to model spatial and contextual dependencies and then optimizes compression parameters using a metaheuristic search with weighted vector quantization. This line of work treats neural networks as feature and context learners while retaining classical quantization and optimization components in the coding loop.

Despite the progress of hybrid designs, several challenges remain, including standardized evaluation protocols, principled ways to balance traditional and learned components, and clinically validated frameworks that maintain diagnostic fidelity across modalities. In particular, hybrid methods would benefit from reporting not only rate--distortion curves but also structure- and task-aware criteria, such as ROI fidelity and downstream segmentation or detection consistency, to better reflect medical imaging requirements.

\color{revblue}
\subsection{Summary and Discussion}
\subsubsection{Qualitative Evolution}
As summarized in Table~\ref{tab:2d_methods}, 2D medical image compression has gradually shifted from handcrafted transform coding toward data-driven neural representations. Traditional methods, exemplified by DCT-based pipelines and wavelet-based standards such as JPEG2000, rely on fixed basis functions and benefit from mature implementations and strong standardization. However, at low bit rates they may suffer from visible artifacts or loss of subtle textures, especially when complex anatomical or pathological patterns are present. In contrast, learning-based approaches, from early CNN-based codecs to more recent Transformer- and INR-based models, provide more flexible feature representations. By optimizing rate--distortion objectives end to end, and in some cases incorporating ROI- or structure-aware constraints, these methods can better preserve diagnostically relevant structures and reduce artifacts that are difficult to handle with fixed transforms.

\subsubsection{Quantitative Performance}
Table~\ref{tab:quant_2D} provides a chronological summary of quantitative results reported in representative studies. Although experimental settings vary across papers (datasets, bit depth, preprocessing, and metric protocols), many recent learning-based codecs report improved reconstruction quality under their respective evaluation settings compared with earlier traditional baselines. For example, a wavelet-domain vector quantization method~\cite{jiang2012medical} reported a PSNR of about 37.5~dB for CT images at a fixed compression ratio (CR=25) under its setting. More recent neural codecs have reported higher PSNR values on their own datasets and protocols; for instance, the segmentation-driven approach in~\cite{vikraman2024segmentation} reported a PSNR around 45.5~dB for brain MRI under a segmentation-guided setting. In addition, some learning-based methods are evaluated under explicitly constrained compression settings (e.g., fixed CR~\cite{sridhar2022optimal} or fixed bpp~\cite{gao2023sinco}), highlighting the importance of reporting rate controls when interpreting PSNR/SSIM improvements.

Despite these reported quantitative gains, deep models typically require substantially more computation than highly optimized standards such as JPEG2000, and their behavior under rare pathological patterns remains a safety-critical concern. Ensuring faithful preservation of subtle diagnostic cues, ideally supported by structure- and task-aware evaluations in addition to PSNR/SSIM, is essential for clinical deployment.
\color{black}

\section{3D/4D medical images compression methods}

Many clinically important medical modalities, such as CT and MRI, are volumetric, including both static 3D volumes and dynamic 4D sequences. With advances in acquisition technology, time-resolved imaging has become increasingly common for capturing the dynamics of tissues and organs. Due to higher dimensionality and often higher bit depth, volumetric data typically impose substantially larger storage and transmission burdens than 2D images, which motivates dedicated compression methods.

Although conventional image and video codecs are effective in many settings, they can be suboptimal for volumetric medical data because they do not always fully exploit inter-slice continuity, spatio-temporal dependencies, and modality-specific characteristics such as anisotropic sampling and clinically important structural regions. This section reviews compression methods designed for volumetric medical images and organizes them into two categories consistent with earlier sections: traditional approaches that extend classical signal processing techniques to higher dimensions, and learning-based approaches that use neural networks to model complex volumetric patterns. We first discuss traditional methods that adapt transform-based, predictive, and motion-compensation techniques to volumetric data, followed by recent learning-based methods that aim to better preserve anatomical structure while improving coding efficiency.

\subsection{Traditional Medical Images Compression Methods}

\subsubsection{Transform-Based Compression Methods}
\rvnew{
Transform-based methods form a fundamental framework for 3D/4D medical image compression. A typical pipeline applies a 3D transform or multiresolution decomposition to exploit intra-slice and inter-slice correlations, followed by optional quantization for lossy coding, coefficient organization across scales or subbands, and entropy coding. The development of this line of work builds on general advances in transform coding, including early DCT formulations and their standardization in JPEG~\cite{DCT1974,wallace1991jpeg}, as well as wavelet-based coding that enabled scalable representations. Wavelet-based compression for volumetric data~\cite{muraki1992approximation} and the JPEG2000 framework~\cite{taubman2002jpeg2000} further motivated volumetric extensions such as JP3D~\cite{bruylants2009jp3d}, which are particularly relevant for medical datasets.}

Here, we review two representative technical directions for volumetric transform coding, wavelet-based methods and DCT-based techniques, focusing on their core ideas, practical implementations, and how they leverage volumetric continuity in 3D/4D medical imaging.

\textbf{Wavelet-based Methods}. Early and influential work by Kim and Pearlman~\cite{kim1999lossless} extended the hierarchical set-partitioning idea of SPIHT from 2D to 3D by constructing 3D spatial-dependence trees for wavelet coefficients. The resulting design preserves the embedded bitstream property and illustrates how inter-slice continuity can be exploited within a wavelet-domain coding framework.

Sanchez et al.~\cite{sanchez2009symmetry} explored the use of anatomical symmetry in the wavelet domain via symmetry-based prediction. This line of work highlights that incorporating anatomical priors can be beneficial for medical data under appropriate acquisition settings, beyond purely generic statistical modeling.

Extensions to dynamic medical imaging were investigated by Bernab{\'e} et al.~\cite{bernabe2000new}, who considered medical sequences that often exhibit gradual changes and structured motion with stringent fidelity requirements. By extending wavelet representations along the temporal dimension, their approach avoids the complexity of explicit motion estimation and reported improved performance over MPEG-2 under their experimental settings.

Hybrid wavelet-domain designs have also been explored, for example by combining multiresolution wavelet representations with vector quantization~\cite{angelidis1994mr,ammah2019robust,linde2003algorithm}. These studies suggest that integrating complementary components can improve rate--distortion behavior for medical imagery in certain scenarios.

For threshold-segmented 3D CT data, V{\v s}peliv{\'c} et al.~\cite{vspelivc2012lossless} proposed a lossless compression pipeline that segments voxels according to selected Hounsfield-unit ranges and encodes the result as two streams. Specifically, the method separates the position information of remaining voxels after segmentation, which is compressed using the JBIG standard, from the corresponding segmented voxel values, which are encoded as a second stream. This design leverages the relationship between Hounsfield values and tissue types to improve coding efficiency while preserving perfect reconstruction.

Finally, Bruylants et al.~\cite{bruylants2015wavelet} provided a systematic synthesis and comparative evaluation of wavelet-based volumetric compression methods, which offers useful guidance on design choices and evaluation practices for subsequent research.

\textbf{DCT-based Methods}.
The success of block-based DCT in JPEG also motivated extensions of DCT-style processing to volumetric medical data. In volumetric settings, DCT-based designs typically follow a block-based workflow by forming 3D cubes (or processing 2D slices with additional inter-slice handling), applying DCT to exploit local correlations, and then performing optional quantization or thresholding and entropy coding for the final bitstream. Application-specific optimizations have been explored for specialized modalities. For example, Xue et al.~\cite{xue20213d} studied DCT-based compression for wireless capsule endoscopy and reported improved performance under their experimental settings through 3D data-structure reconstruction and computational optimizations tailored to the correlation patterns of endoscopy data.

\textbf{Other transform-based Methods}. To address computational and memory constraints in volumetric transform coding, Senapati et al.~\cite{senapati2016volumetric} proposed listless coding strategies that reduce auxiliary data structures required by conventional SPIHT-style implementations, reporting substantial memory savings compared with direct extensions. Their work also explored hybrid transform designs that combine wavelet transforms in spatial dimensions with Karhunen--Lo{\`e}ve transforms in spectral dimensions, illustrating how multiple transforms can be integrated to better match the structure of specific volumetric data.

\subsubsection{Prediction and Motion Compensation-based Methods}
\rvnew{
Prediction and motion-compensation-based methods target dynamic 4D medical imaging by exploiting spatio-temporal redundancy across consecutive frames or volumes. A typical pipeline follows the standard video-coding structure: intra-frame coding for spatial redundancy, inter-frame prediction with motion estimation/compensation, residual coding, and entropy coding, with variants adapted to medical constraints such as high bit depth, strict lossless or near-lossless requirements, and structured anatomical motion.}

Sanchez et al.~\cite{sanchez2008efficient} proposed an early motion-compensated lossless compression method for dynamic volumetric medical data. Their work leverages motion information between consecutive frames to improve compression efficiency under their reported settings and illustrates how motion-compensated prediction can be adapted to volumetric medical sequences. Relatedly, Sanchez et al.~\cite{sanchez2006lossless} investigated applying the H.264/AVC framework to lossless compression of 4D medical images, showing that standardized prediction and motion-compensation mechanisms in video codecs can serve as practical baselines for handling spatio-temporal redundancy in medical data.

Beyond single-view temporal coding, Martin et al.~\cite{martin2008analysis} analyzed the feasibility of using multi-view video coding (MVC) concepts for volumetric medical imagery, aiming to exploit correlation across multiple related image streams when such acquisitions are available. This line of work further supports the idea that standardized video-coding tools can be transferred to medical imaging when the acquisition geometry and correlation structure are compatible.

Hybrid designs have also been explored. Kassim et al.~\cite{kassim2005motion} combined motion-compensated prediction with integer wavelet transforms, integrating temporal-domain motion compensation with spatial-domain multiresolution representation and supporting progressive reconstruction from lossy to lossless under their design.

More recent efforts have focused on newer video-coding standards. Guarda et al.~\cite{guarda2017method} proposed improvements to HEVC lossless coding for volumetric medical images by tuning coding tools to medical-image characteristics and reported performance gains under their evaluation settings. Parikh et al.~\cite{parikh2017high} studied high bit-depth medical image compression using HEVC-based designs. Overall, these works suggest that modern video coding standards provide a strong and workflow-compatible foundation for 3D/4D medical image compression, especially when combined with medical-specific constraints and evaluation protocols.

\subsubsection{Predictive Coding Methods}
Unlike transform-based approaches, predictive coding exploits spatial correlations by predicting voxel values from previously coded neighbors and encoding only the prediction residuals. \rvnew{A typical volumetric predictive pipeline consists of 3D neighborhood prediction, residual computation, and context-based entropy coding, optionally combined with adaptive partitioning to better match local anatomical structures.}

Lucas et al.~\cite{lucas2017lossless} presented a 3D median root prediction (3D-MRP) framework that adapts MRP-style predictors to volumetric medical images. The method exploits correlations across the three spatial dimensions and employs 3D block octree partitioning to enable adaptive processing based on local volumetric characteristics, aiming to improve coding efficiency under lossless reconstruction constraints.

Santos et al.~\cite{santos2015contributions,santos2016compression} further enhanced the MRP framework by incorporating several complementary components. They introduced bi-directional prediction to exploit continuity across slices when applicable~\cite{santos2015contributions}, designed predictors that adapt to local intensity variations and boundary transitions in medical volumes~\cite{santos2016compression}, and applied histogram packing techniques to better utilize the statistical structure of prediction residuals~\cite{santos2016compression}. These additions were reported to improve lossless compression performance under their evaluation settings.

In parallel, alternative prediction frameworks were investigated. D{\'\i}ez et al.~\cite{diez2005lossless} proposed an LSE-3D approach that uses least-squares estimation tailored for volumetric medical data. The authors reported competitive compression performance under their experimental protocol while maintaining lossless reconstruction, and the method offers a relatively direct computational formulation through least-squares prediction.

More recently, geometry- and anatomy-aware predictive coding has been explored. Song et al.~\cite{song2018lossless} proposed anatomy-aware partitioning and adaptive selection of prediction operators guided by localized boundary cues and continuity modeling, aiming to better preserve diagnostically important structural gradients. This line of work suggests that incorporating anatomical priors into predictive modeling can improve volumetric predictive coding when acquisition characteristics and clinical constraints are properly considered.

\subsection{Learning based medical images compression methods}

Learning-based approaches have become increasingly important for 3D medical image compression, leveraging data-driven representation learning to model volumetric structures and improve rate--distortion behavior~\cite{bourai2024deep}. \textcolor{revblue}{Overall, their evolution largely mirrors the trends discussed for 2D learning-based codecs in Section~II-B1), so we focus here on how these ideas are organized and instantiated for volumetric data.}

\rvnew{
From the perspective of learning objectives and what is stored as the compressed representation, existing methods can be grouped into two complementary paradigms. The first category, Explicit Feature Learning Methods, learns an end-to-end or modular neural codec that outputs a bitstream for storage and transmission. In many designs, a learned analysis transform maps a 3D volume (or a sequence of slices) to latent features, followed by quantization and entropy modeling with context/hyperprior networks, and a synthesis transform reconstructs the volume from the decoded latents. Learning-based modules may target specific stages of a traditional pipeline such as prediction, transform design, quantization, and entropy coding, often using 3D convolutions or inter-slice context models to exploit volumetric continuity.}

The second category, Implicit Neural Representation Methods, represents volumetric data as a continuous coordinate-to-intensity function parameterized by a neural network. In this case, the compressed form is primarily the network parameters rather than a conventional pixel/voxel bitstream. Specifically, an INR model learns a mapping from spatial coordinates (x, y, z) to voxel intensities, optionally with positional encoding, and can reconstruct values at arbitrary coordinates on demand~\cite{sitzmann2020implicit}.

This distinction reflects two philosophies of applying deep learning to medical image compression: improving an explicit coding pipeline versus learning a compact volumetric representation. Both directions have reported favorable results compared with conventional baselines under their respective evaluation settings. For example, some explicit codecs incorporate structure- or ROI-aware constraints to better preserve diagnostically important regions~\cite{min2022lossless}, while INR-based methods exploit continuous function approximation to achieve compact representations for volumetric data~\cite{sitzmann2020implicit}. In the following, we review these two paradigms with an emphasis on their volumetric instantiations, key design choices, and practical considerations.

\definecolor{headerbg}{RGB}{0, 50, 100}         
\definecolor{col_sidebar}{RGB}{235, 242, 250}   
\definecolor{row_odd}{RGB}{255, 255, 255}       
\definecolor{row_even}{RGB}{248, 250, 251}      

\setlist[tabitemize]{
    label=\textbullet,
    leftmargin=*,
    nosep,
    topsep=2pt,         
    itemsep=2pt,        
    after=\vspace{2pt}  
}

\begin{table*}[!t]
\centering
\caption{A comprehensive overview of explicit feature learning methods for 3D/4D medical image compression. We systematically examine the evolution of various compression techniques, tracing the technological progression from conventional CNN-based approaches to state-of-the-art architectures such as Transformers.}
\label{tab_3Ddeep}

\renewcommand{\arraystretch}{1.3} 
\setlength{\tabcolsep}{5pt}        
\setlength{\aboverulesep}{0pt}
\setlength{\belowrulesep}{0pt}
\arrayrulecolor{headerbg}

\begin{tabular}{
    >{\columncolor{col_sidebar}\bfseries}c  
    c                                       
    c                                       
    M{2.2cm}                                
    M{1.5cm}                                
    M{7.5cm}                                
}

\toprule

\rowcolor{headerbg} 
\head{Ref.} & 
\head{Year} & 
\head{Type} & 
\head{Method} & 
\head{Modality} & 
\head{Key Contributions \& Limitations} \\

\rowcolor{row_even}
\cite{luu2021efficiently} & 2021 & Lossy & 
CNN + \newline Anisotropic Diff. & 
CT & 
\begin{tabitemize}
    \item Uses CNN to extract organ probability maps for efficiency.
    \item Combines anisotropic diffusion to preserve features.
    \item \textit{Limitation:} Struggles with fine high-frequency details at boundaries.
\end{tabitemize} \\

\rowcolor{row_odd}
\cite{nagoor2021medzip} & 2021 & Lossless & 
LSTM & 
CT, MRI & 
\begin{tabitemize}
    \item 3D sequence prediction model for voxel dependencies.
    \item Effective for long-range dependencies in 3D data.
    \item \textit{Limitation:} High computational cost; scalability issues.
\end{tabitemize} \\

\rowcolor{row_even}
\cite{min2022lossless} & 2022 & Lossless & 
CNN & 
CT & 
\begin{tabitemize}
    \item Divides images into anatomical regions for adaptive prediction.
    \item Preserves anatomical structure.
    \item \textit{Limitation:} May not generalize well across diverse regions.
\end{tabitemize} \\

\rowcolor{row_odd}
\cite{xue2022aiwave} & 2022 & Both & 
CNN + \newline Wavelet Trans. & 
CT, MRI & 
\begin{tabitemize}
    \item Introduces 3D trained wavelet-like transform.
    \item Effective for both lossless and lossy compression.
    \item \textit{Limitation:} May introduce artifacts at lower compression ratios.
\end{tabitemize} \\

\rowcolor{row_even}
\cite{chen2022exploiting} & 2022 & Lossless & 
CNN + \newline Gating Mech. & 
MRI & 
\begin{tabitemize}
    \item Uses hierarchical intra-slice auxiliary features.
    \item Combines multi-scale intra- and inter-slice info.
    \item \textit{Limitation:} Complex structure increases overhead.
\end{tabitemize} \\

\rowcolor{row_odd}
\cite{nagoor2022sampling} & 2022 & Lossless & 
LSTM & 
CT, MRI & 
\begin{tabitemize}
    \item Explores input configurations and sampling schemes.
    \item Robust performance across configurations.
    \item \textit{Limitation:} Long training times and large data needs.
\end{tabitemize} \\

\rowcolor{row_even}
\cite{chen2023streaming} & 2023 & Lossless & 
CNN + \newline Gated RNN & 
CT, MRI & 
\begin{tabitemize}
    \item Captures inter-slice dependencies with gated RNN.
    \item Enhances reconstruction quality with recurrent CNN.
    \item \textit{Limitation:} Limited scalability for very large datasets.
\end{tabitemize} \\

\rowcolor{row_odd}
\cite{liu2024bilateral} & 2024 & Lossless & 
CNN + \newline Attention & 
CT, MRI & 
\begin{tabitemize}
    \item Bilateral context modeling for anatomical structure.
    \item First architecture aware of anatomical symmetry.
    \item \textit{Limitation:} Requires extensive training resources.
\end{tabitemize} \\

\rowcolor{row_even}
\cite{wang2024learning} & 2024 & Lossless & 
Transformer & 
CT, MRI, EM & 
\begin{tabitemize}
    \item Divides high bit-depth volumes into significant bits.
    \item Hybrid of traditional codecs and transformer models.
    \item \textit{Limitation:} Increased complexity in implementation.
\end{tabitemize} \\

\bottomrule

\end{tabular}
\end{table*}

\subsubsection{Explicit Feature Learning Methods} 

Deep learning-based 3D medical image compression has been explored with different neural architectures to better model volumetric structure and improve coding efficiency. As summarized in Table~\ref{tab_3Ddeep}, representative designs can be grouped into CNN-based methods that focus on spatial/context modeling, RNN-based approaches that process slices or frames sequentially, and attention-based methods that capture longer-range dependencies.

\textbf{CNN-based methods.} CNN-based codecs replace or augment hand-crafted transforms with learned analysis/synthesis networks and learned entropy models. A typical pipeline maps a 3D volume (or a stack of slices) to latent features using 3D convolutions, applies quantization, encodes the latents with context/hyperprior-based entropy modeling into a bitstream, and reconstructs the volume via a learned decoder. Compared with applying generic transforms uniformly, CNN-based designs can be trained to better match volumetric medical statistics, including homogeneous tissue regions, sharp anatomical boundaries, and modality-dependent noise characteristics, and some works further introduce structure- or ROI-aware constraints to preserve diagnostically important details.

Early studies emphasized anatomy- or ROI-aware processing. Min et al.~\cite{min2022lossless} proposed a lossless framework that partitions regions using anatomical cues and trains region-specific predictors in a divide-and-conquer manner. Luu et al.~\cite{luu2021efficiently} introduced DLAD, combining CNN-based organ segmentation with anisotropic diffusion filtering to guide selective smoothing while preserving key anatomical structures for teleintervention scenarios.

Subsequent work increasingly integrated learned modules into established volumetric coding frameworks. Xue et al.~\cite{xue2022aiwave} proposed aiWave, which replaces the prediction and update filters in a lifting-based wavelet transform with 3D CNNs and incorporates adaptive affine mapping to enable signal-dependent, non-separable transforms; they reported BD-PSNR gains over HEVC under their evaluation protocol. Chen et al.~\cite{chen2022exploiting} developed ICEC to jointly model intra- and inter-slice redundancies via hierarchical compression with Intra-Gate and Inter-Gate mechanisms, reporting improved performance compared with JP3D and HEVC baselines.

Overall, CNN-based volumetric codecs have evolved from anatomy/ROI-aware designs toward stronger context utilization and learned coding components tailored to 3D medical data. Recent lossless volumetric frameworks further exploit long-range and symmetry-aware context via attention-style mechanisms, which we discuss in the attention-based category.

\begin{figure*}[!t]
	\centering
	\includegraphics [width=0.98\textwidth]{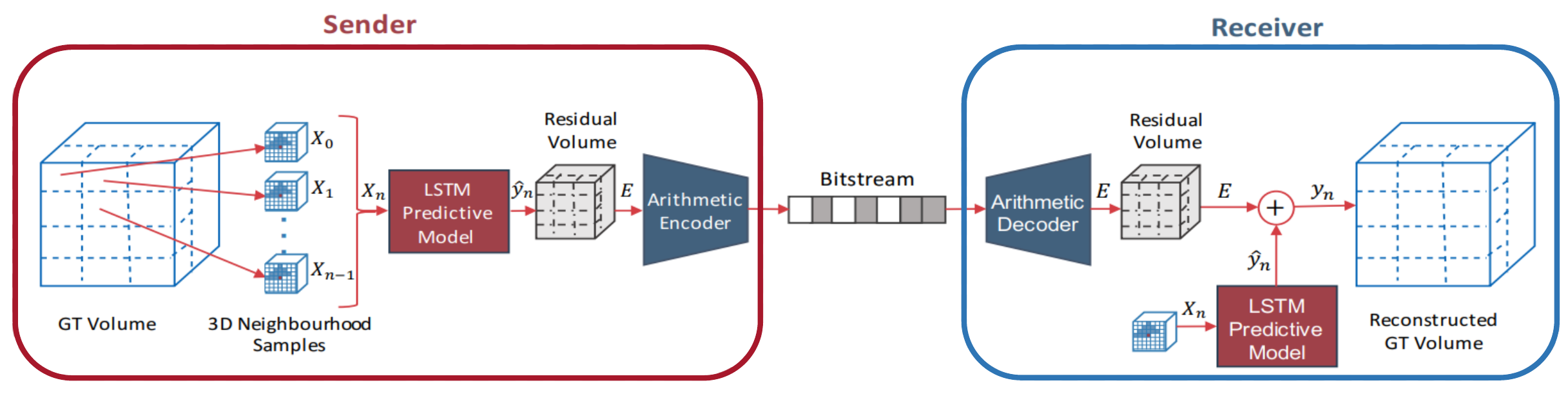}\\
	\caption{\textbf{3D RNN based medical image compression.} Medzip \cite{nagoor2021medzip} uses LSTM neural networks to predict voxel values and achieves compression by encoding prediction residuals.}
	\label{fig:3D_RNNbased}
    \vspace{-0.5cm}
\end{figure*}

\textbf{Recurrent neural network(RNN)-based methods.}
While CNN-based methods primarily model spatial correlations within and across local 3D neighborhoods, volumetric medical data also exhibit strong dependencies along the slice direction. This motivates RNN-based designs that treat a volume as a sequence and use recurrent memory to model inter-slice continuity, so that only prediction residuals need to be coded.

Nagoor et al.~\cite{nagoor2021medzip} proposed MedZip, which integrates an LSTM-based sequence predictor with arithmetic coding of residuals. As illustrated in Fig.~\ref{fig:3D_RNNbased}, the framework partitions 3D or 4D medical data into sequential segments, predicts target samples using an LSTM model, entropy-codes the resulting residuals into a bitstream, and reconstructs the original data via arithmetic decoding with the same predictor. The authors reported improved lossless compression performance compared with JPEG-LS, JPEG2000, JP3D, HEVC, and PPMd under their evaluation protocol.

The same group further investigated input configurations and sampling schemes for the many-to-one LSTM predictor in MedZip~\cite{nagoor2022sampling}. Using experiments with cross-validation on CT datasets, they reported that a pyramid-shaped sampling strategy provides a favorable trade-off between compression effectiveness and model input design, yielding additional gains over several well-established lossless baselines under their settings.

More parameter-efficient recurrent architectures have also been explored. Chen et al.~\cite{chen2023streaming} introduced a gated recurrent convolutional network that combines convolutional structures with fusion-gate mechanisms to capture inter-slice dependencies in a streaming-friendly manner. They reported comparable compression effectiveness while using substantially fewer parameters than other learning-based frameworks, which can reduce deployment overhead.

Overall, RNN-based volumetric codecs improve compression by explicit sequence modeling along the slice or time dimension, but their sequential dependency can reduce parallelism and increase latency compared with fully parallel 3D CNN or attention-based designs.

\textbf{Attention-based methods.}
CNN-based codecs are effective at local feature extraction but are limited by local receptive fields, while RNN-based designs can model inter-slice continuity but often introduce sequential processing bottlenecks. Attention mechanisms and Transformer architectures provide a complementary solution by explicitly modeling long-range dependencies and global context in volumetric data, which can strengthen correlation modeling across slices and improve entropy prediction for compression.

Liu et al.~\cite{liu2024bilateral} proposed BCM-Net for residual coding in lossless 3D medical image compression by exploiting bilateral correlations commonly observed in anatomical volumes. The method designs two context-extraction modules: SICE captures bilateral intra-slice correlations guided by anatomical symmetry using attention-based correlation mining, and BICE explores bi-directional inter-slice references using cross-attention style context fusion. The overall design follows a two-layer residual-coding framework, where a VVC-coded base layer provides a strong prediction and the residuals are then coded losslessly using BCM-Net, enabling exact reconstruction when both the base layer and residual bitstreams are available.

\begin{figure*}[!t]
	\centering
	\includegraphics [width=0.98\textwidth]{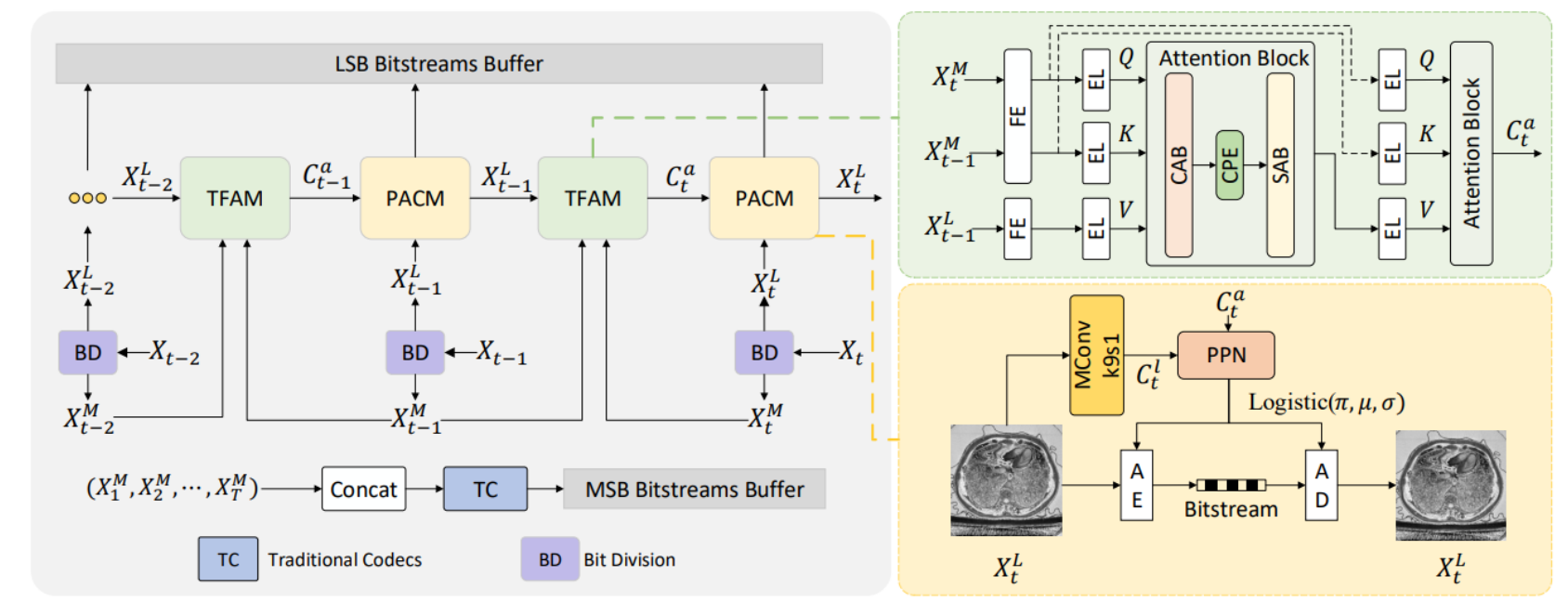}\\
	\caption{Representative Architecture of Attention-based 3D Medical Image Compression: BD-LVIC Framework \cite{wang2024learning}.}
	\label{fig:bd_lvic_architecture}
    \vspace{-0.5cm}
\end{figure*}

Wang et al.~\cite{wang2024learning} presented BD-LVIC, a lossless framework tailored for high bit-depth volumetric medical data. The method divides a volume into a most-significant-bit volume (MSBV) and a least-significant-bit volume (LSBV) and applies a dual-stream coding strategy, with attention-based feature alignment and context modeling focusing on the detailed information in the LSBV. As illustrated in Fig.~\ref{fig:bd_lvic_architecture}, the architecture integrates Transformer-style long-range modeling with slice-aware context modules to jointly exploit intra-slice and inter-slice redundancies. These attention-based volumetric codecs can improve lossless coding efficiency under their reported settings, but they typically require higher computation and memory than conventional volumetric standards.

\subsubsection{Implicit Neural Representation Methods}
INR-based volumetric compression has been explored along multiple directions. In this survey, we group representative methods into four categories, organized by whether the main contribution lies in the representation/architecture design or in the optimization priors: (i) hybrid representation architectures, (ii) semantic-aware compression, (iii) frequency-aware optimization, and (iv) structured or multi-scale representations.

\textbf{Hybrid representation architectures.}
Pure INR approaches based on a single MLP can face scalability and computational challenges on high-dimensional 3D/4D data~\cite{dupont2021coin,sitzmann2020implicit}. To mitigate these issues, hybrid architectures combine continuous INR representations with discrete structures, decompositions, or classical modules to improve efficiency~\cite{zheng2024hybrid,sheibanifard2023novel}. Zheng et al.~\cite{zheng2024hybrid} introduced a mixed-representation framework that partitions the representation space to better handle the spatio-temporal complexity of 4D medical datasets. For volumetric data, Sheibanifard and Yu~\cite{sheibanifard2023novel} proposed a three-stage design that couples downsampling with a SIREN-based INR and a learned reconstruction module. More broadly, ROI-aware hybrid pipelines that combine classical transforms and neural components have also been studied for medical compression~\cite{li2025towards}, and can be viewed as complementary to INR-based designs when targeting efficiency and workflow compatibility. Overall, these hybrid strategies aim to improve scalability and reduce overhead while maintaining reconstruction fidelity under their reported settings.

\textbf{Semantic-aware compression methods.}
Semantic-aware INR compression introduces task- or structure-related priors so that coding resources are preferentially allocated to diagnostically or scientifically important content. Ma et al.~\cite{ma2024semantic} explored saliency-guided INR compression for biomedical imagery by incorporating semantic cues into the representation learning process, reporting very high compression ratios under their protocol while preserving application-relevant fidelity. UniCompress~\cite{yang2024unicompress} introduced a distillation-based framework that incorporates priors (e.g., frequency-domain codebooks and wavelet components) to strengthen INR representational power. For specialized biological imaging, the Implicit Neural Image Field approach~\cite{dai2024implicit} tailors the INR design and training objectives to domain-specific redundancies to preserve microscopic structures that are relevant for scientific analysis.

\textbf{Frequency-aware optimization methods.}
Conventional spatial-domain INRs can exhibit spectral bias, which may limit their ability to represent high-frequency components efficiently. Frequency-aware methods address this by explicitly shaping the representation or parameter allocation according to spectral characteristics. Yang et al.~\cite{yang2023sci} proposed Spectrum Concentrated Implicit Neural Compression (SCI), which partitions biomedical data into spectrum-concentrated blocks and applies parameter-efficient INR models to each block, reporting improved compression performance over spatial-only INR baselines under their evaluated datasets.

\textbf{Structured representation methods.}
Structured INR compression introduces hierarchical or multi-scale representations to improve parameter efficiency. Str\"{u}mpler et al.~\cite{strumpler2022implicit} presented an INR compression pipeline with meta-learned initialization, quantization-aware training, and entropy coding, where the quantized network parameters form the compressed representation. Building on this direction, hierarchical decomposition has been explored to better match the multi-scale nature of medical imaging data~\cite{yang2023tinc2}, offering an alternative to monolithic INR models and enabling more efficient parameter utilization.

\definecolor{headerbg}{RGB}{0, 50, 100}         
\definecolor{col_sidebar}{RGB}{235, 242, 250}   
\definecolor{row_odd}{RGB}{255, 255, 255}       
\definecolor{row_even}{RGB}{248, 250, 251}      

\setlist[tabitemize]{
    label=\textbullet,
    leftmargin=*,
    nosep,
    topsep=3pt,
    itemsep=2pt,
    after=\vspace{2pt}
}

\begin{table*}[t]
\centering
\caption{Qualitative Comparison of Representative 3D/4D Medical Image Compression Methods}
\label{tab:3d_methods}

\renewcommand{\arraystretch}{1.35} 
\setlength{\tabcolsep}{5pt}        
\setlength{\aboverulesep}{0pt}
\setlength{\belowrulesep}{0pt}
\arrayrulecolor{headerbg}

\begin{tabular}{
    >{\columncolor{col_sidebar}\bfseries}M{2.8cm} 
    M{3.2cm}   
    M{3.4cm}   
    M{3.4cm}   
    M{3.4cm}   
}

\toprule

\rowcolor{headerbg} 
\head{Category} & 
\head{Representative Works} & 
\head{Key Characteristics} & 
\head{Strengths} & 
\head{Limitations} \\

\rowcolor{row_even}
Transform-based Compression &
Wavelet for 3D~\cite{kim1999lossless}; \newline
3D DCT Volumetric~\cite{raj2007novel}; \newline
Wavelet/KL Hybrid~\cite{senapati2016volumetric} &
\begin{tabitemize}
    \item Multi-resolution decomposition (DWT/DCT)
    \item Handles volumetric data
    \item Exploits inter-slice correlations
\end{tabitemize} &
\begin{tabitemize}
    \item Good R-D performance
    \item Effective spatial correlation use
    \item Near-lossless control
\end{tabitemize} &
\begin{tabitemize}
    \item High complexity with dimensions
    \item Blocking artifacts
    \item Limited temporal capture (4D)
\end{tabitemize} \\

\rowcolor{row_odd}
Prediction \& Motion Compensation &
Motion-Comp Lossless~\cite{sanchez2008efficient}; \newline
HEVC-based 3D~\cite{guarda2017method}; \newline
High Bit-depth~\cite{parikh2017high} &
\begin{tabitemize}
    \item Exploits temporal redundancies
    \item Predicts subsequent frames/slices
    \item Focus on dynamic imaging
\end{tabitemize} &
\begin{tabitemize}
    \item Low computational cost
    \item Efficient for temporal data
    \item Preserves dynamic changes
\end{tabitemize} &
\begin{tabitemize}
    \item Sensitive to motion artifacts
    \item Struggles with complex variations
    \item Frame-to-frame dependency
\end{tabitemize} \\

\rowcolor{row_even}
Predictive Coding &
3D Median Root Pred.~\cite{lucas2017lossless}; \newline
Enhanced MRP~\cite{santos2015contributions,santos2016compression} &
\begin{tabitemize}
    \item Pixel/Residual prediction in slices
    \item Tailored for lossless/near-lossless
    \item Simple entropy coding
\end{tabitemize} &
\begin{tabitemize}
    \item Simple architecture
    \item Low memory/compute overhead
    \item Excellent fine detail preservation
\end{tabitemize} &
\begin{tabitemize}
    \item Limited for complex textures
    \item Weak on long-range correlations
    \item Lower efficiency than DL
\end{tabitemize} \\

\rowcolor{row_odd}
Explicit Feature Learning &
Anatomical DL~\cite{min2022lossless}; \newline
LSTM-based~\cite{nagoor2021medzip}; \newline
Bilateral Context~\cite{liu2024bilateral} &
\begin{tabitemize}
    \item DL-based feature extraction
    \item Focus on anatomical structure
    \item Better compression performance
\end{tabitemize} &
\begin{tabitemize}
    \item Adapts to specific modalities
    \item Handles complex anatomy well
    \item High compression ratios
\end{tabitemize} &
\begin{tabitemize}
    \item Data-intensive training
    \item Large annotated datasets needed
    \item Prone to overfitting
\end{tabitemize} \\

\rowcolor{row_even}
Implicit Neural Representation &
Hybrid INR~\cite{zheng2024hybrid}; \newline
Semantic INR~\cite{ma2024semantic} &
\begin{tabitemize}
    \item Continuous coord-to-intensity
    \item Uses positional encodings
    \item Network parameterizes volume
\end{tabitemize} &
\begin{tabitemize}
    \item Highly compact
    \item Sub-pixel accuracy
    \item Resolution independence
    \item Ideal for high-res volumes
\end{tabitemize} &
\begin{tabitemize}
    \item Slow training
    \item Computationally demanding
    \item Scaling challenges (large data)
\end{tabitemize} \\

\bottomrule

\end{tabular}
\end{table*}

\definecolor{headerbg}{RGB}{0, 50, 100}
\definecolor{col_sidebar}{RGB}{235, 242, 250}
\definecolor{row_odd}{RGB}{255, 255, 255}
\definecolor{row_even}{RGB}{248, 250, 251}

\begin{table*}[t]
\centering
\caption{Quantitative comparison of representative 3D/4D medical image compression methods (values reported from original papers).
The ``Dataset / Experimental Setting'' column summarizes the data source (DS), dimensionality, and bit-depth (BD), together with key characteristics explicitly stated in the original works.
Note that experimental conditions (datasets, bit-depths, evaluation protocols, and compression modes) vary across references; therefore, the comparison is indicative rather than a strictly controlled benchmark.
Reported runtime values are listed as-is from the original papers and should not be interpreted as a fair efficiency comparison.}
\label{tab:quant_3D}
\footnotesize

\renewcommand{\arraystretch}{1.15}
\setlength{\tabcolsep}{4pt}
\arrayrulecolor{headerbg}

\begin{tabular}{
    >{\columncolor{col_sidebar}\bfseries}c
    c
    l
    l
    >{\raggedright\arraybackslash}p{5.5cm}
    c
    c
    c
    c
    c
}

\toprule

\rowcolor{headerbg}
& & & & & & & & \multicolumn{2}{c}{\textbf{\textcolor{white}{Reported Runtime (s)}}} \\

\rowcolor{headerbg}
\multirow{-2}{*}{\textbf{\textcolor{white}{Ref.}}} &
\multirow{-2}{*}{\textbf{\textcolor{white}{Year}}} &
\multirow{-2}{*}{\textbf{\textcolor{white}{Type}}} &
\multirow{-2}{*}{\textbf{\textcolor{white}{Method}}} &
\multirow{-2}{*}{\textbf{\textcolor{white}{Dataset / Experimental Setting}}} &
\multirow{-2}{*}{\textbf{\textcolor{white}{Bpp}}} &
\multirow{-2}{*}{\textbf{\textcolor{white}{PSNR}}} &
\multirow{-2}{*}{\textbf{\textcolor{white}{SSIM}}} &
\head{Comp.} &
\head{Decomp.} \\

\rowcolor{row_even}
\cite{kim1999lossless} & 1999 & Lossless & 3D SPIHT
& \makecell[l]{DS=MR (chest, liver, head) + CT skull\\ BD=8b, $256\times256$ slices, GOS=16}
& 1.96 & -- & -- & -- & -- \\

\rowcolor{row_odd}
\cite{santos2015contributions} & 2015 & Lossless & MRP Inter Diff
& \makecell[l]{DS=4 CT + 4 MRI volumes\\ BD=8b, $256\times256$ slices, 58--203 slices}
& 1.04 & -- & -- & -- & -- \\

\rowcolor{row_even}
\cite{senapati2016volumetric} & 2016 & Lossy & 3D-HLCK
& \makecell[l]{DS=Brain MRI, DICOM knee, angiogram\\ BD=NS, $128\times128$ slices (8 slices)}
& 0.50 & 25.26 & -- & -- & -- \\

\rowcolor{row_odd}
\cite{lucas2017lossless} & 2017 & Lossless & 3D-MRP
& \makecell[l]{DS=Stanford VDA, TCIA (CT/MRI volumes)\\ BD=8/16b, $\sim256\times256$ slices, 58--203 slices}
& 0.89 & -- & -- & 15578.53 & 2.04 \\

\rowcolor{row_even}
\cite{luu2021efficiently} & 2021 & ROI Lossless & CNN + AD
& \makecell[l]{DS=Abdominal CT\\ BD=16b, 3D volumes (21--343 slices)}
& 1.24 & 70.03 & 0.95 & 138.5 & 21.6 \\

\rowcolor{row_odd}
\cite{chen2022exploiting} & 2022 & Lossless & CNN + ICEC
& \makecell[l]{DS=MRNet, Brain MRI\\ BD=8--12b, 3D volumes}
& 3.84 & -- & -- & 1.16 & -- \\

\rowcolor{row_even}
\cite{nagoor2022sampling} & 2022 & Lossless & LSTM
& \makecell[l]{DS=CT torso (private), MRI head/neck\\ BD=16b, high-resolution 3D volumes}
& 4.51 & -- & -- & 4.99 & 2004 \\

\rowcolor{row_odd}
\cite{chen2023streaming} & 2023 & Lossless & CNN + Gated RNN
& \makecell[l]{DS=MRNet, CHAOS, DeepLesion\\ BD=8--12b, 3D volumes}
& 3.71 & -- & -- & 0.09 & -- \\

\rowcolor{row_even}
\cite{xue2022aiwave} & 2023 & Lossless & CNN + Wavelet
& \makecell[l]{DS=FAFB, FIB-25, Chaos-CT, MRI-heart\\ BD=8--32b (CT/MRI$\rightarrow$16b)}
& 4.61 & -- & -- & 986.97 & 982.03 \\

\rowcolor{row_odd}
\cite{liu2024bilateral} & 2024 & Lossless & CNN + BCM-Net &
\makecell[l]{DS=MRNet, MosMedData, TRABIT\\ BD=8--16b (MRI/CT volumes)} &
4.41 & -- & -- & 179.3 & 4.75 \\

\rowcolor{row_even}
\cite{wang2024learning} & 2024 & Lossless & Transformer &
\makecell[l]{DS=MRNet, Chaos-CT, Covid-CT, Trabit-MRI\\ BD=8--16b volumetric data} &
3.46 & -- & -- & 0.65 & 0.65 \\

\rowcolor{row_odd}
\cite{zheng2024hybrid} & 2024 & Lossy & INR + Discrete &
\makecell[l]{DS=ACDC cardiac MRI (4D), openfMRI\\ BD=$>$8b (high bit-depth MRI)} &
0.054 & 52.97 & -- & 228.8 & 0.61 \\

\rowcolor{row_even}
\cite{ma2024semantic} & 2024 & Lossy & Semantic INR &
\makecell[l]{DS=ACDC cardiac MRI (4D),\\ BD=$>$8b (high dynamic range)} &
0.01 & -- & -- & -- & -- \\

\bottomrule
\end{tabular}
\end{table*}

\color{revblue}

\subsection{Summary and Discussion}
\subsubsection{Qualitative Evolution}
As summarized in Table~\ref{tab:3d_methods}, progress in 3D/4D medical image compression has largely been driven by increasingly effective exploitation of inter-slice and temporal redundancies. Early solutions extended 2D image/video coding tools to volumetric settings, for example, JP3D for 3D wavelet-based coding~\cite{bruylants2009jp3d} and video codecs such as H.264/AVC for dynamic sequences, often relying on block-based transforms and motion-compensated prediction. While effective, these approaches can be computationally demanding and may introduce blocking or cubic artifacts in block-based configurations. More recently, learning-based codecs, including 3D CNNs and recurrent models, have enabled stronger volumetric context modeling by explicitly capturing continuity along the slice (Z) dimension. In parallel, INR-based approaches have attracted growing interest by representing a volume as a continuous coordinate-to-intensity function rather than a discrete voxel grid, which naturally supports resolution-independent queries and can be beneficial for workflows involving multi-planar reformatting.

\subsubsection{Quantitative Performance}
Table~\ref{tab:quant_3D} highlights the gains obtained by moving from slice-wise processing toward explicit 3D/4D context modeling. By jointly modeling inter-slice and temporal correlations, recent learning-based frameworks can reduce redundancy more effectively than applying 2D codecs independently to each slice. For example, the transformer-based BD-LVIC framework~\cite{wang2024learning} compares against strong traditional codecs (e.g., JPEG-XL and HEVC-RExt intra) and reports state-of-the-art lossless BPV, surpassing JPEG-XL by 11.7\% on their reported datasets/protocol. Beyond lossless settings, several INR-based studies suggest that continuous implicit representations can be competitive for certain dynamic or scientific imaging scenarios at very high compression ratios, although the exact operating points and bpp-equivalents depend on the data definition, bit-depth, and protocol used in each work.

Despite the performance improvements shown in Table~\ref{tab:quant_3D}, volumetric processing can be substantially more resource-intensive than 2D coding. Memory and computation typically scale with the volume size, and large 3D studies can impose significant GPU/CPU demands. Moreover, while INRs can store a compact set of parameters, their per-subject optimization (encoding) time can be a bottleneck for time-sensitive clinical workflows compared with highly optimized conventional codecs.

\color{revblue}

\section{Critical Issues and Practical Challenges}
While the preceding sections summarized the evolution of traditional coding, learning-based approaches, and emerging paradigms such as INRs, here we provide a critical analysis along four dimensions: (1) the continued dominance of transform and predictive coding in medical standards; (2) limitations that hinder clinical adoption of learning-based approaches including VAE, GAN, and INR models; (3) challenges in evaluating algorithms within clinical workflows; and (4) practical obstacles in transitioning research prototypes to DICOM and PACS environments.

\subsection{Why Transform-Based Methods Still Dominate Medical Standards}
Although deep learning has produced impressive advances in lossy and lossless compression for natural images~\cite{liu2022learning,moorthy2021deep}, medical standards such as JPEG-LS, JPEG2000, and JP3D remain rooted in well-standardized transform and predictive coding frameworks~\cite{pinho2017current,liu2017current}. This persistence is not accidental. Clinical imaging imposes strict requirements on traceability, reproducibility, interpretability, and long-term archival stability, which favor mathematically transparent transforms and well-defined codestream syntax.

First, transform/predictive standards exhibit stable and predictable behavior across diverse modalities including CT, MRI, and ultrasound. In contrast, learning-based codecs may require modality-dependent training, calibration, or careful protocol control for robust deployment, which complicates standardization. Second, standardized codestream syntax enables deterministic decoding and consistent reconstruction from the same bitstream across platforms and software versions, supporting auditability and long-term archival use. Learning-based systems, especially those relying on stochastic training, regularization, or generative components, can be harder to constrain and validate for reproducible behavior unless carefully engineered. Third, traditional codecs benefit from decades of ecosystem support: they are integrated into DICOM Part~5~\cite{dicom2021part5} and widely supported by PACS vendors, whereas deep-learning codecs still lack broadly adopted standard codestream definitions and interoperability guarantees.

\subsection{Why INR, VAE, and GAN-Based Methods Remain Experimental}
Despite the strong promise of INRs~\cite{sitzmann2020implicit} and generative compression models such as VAE- or GAN-based architectures~\cite{mishra2022deep}, their clinical deployment remains limited.

First, these models rely on implicit priors learned from data. While such priors can produce visually convincing reconstructions at high compression ratios, they raise concerns about hallucinated structures, loss of subtle diagnostic cues, and structure-preserving inconsistencies in reconstructed CT or MRI volumes. In contrast to many natural-image applications where perceptual plausibility is often acceptable, medical imaging requires strict preservation of low-contrast lesions, micro-calcifications, vessel boundaries, and subtle tissue textures. INR-based reconstructions may oversmooth fine details or bias structures toward the learned prior, which is difficult to validate clinically.

Second, the encoding cost of INR-based approaches remains substantially higher than that of conventional codecs. Even optimized variants such as COIN++~\cite{dupont2022coin++} or SINCO~\cite{gao2023sinco} often require per-sample fitting or heavy optimization, which is incompatible with clinical-scale workloads that can span very large multi-study archives.

Third, generative codecs require comprehensive testing across heterogeneous scanners, protocols, reconstruction kernels, and acquisition conditions. Robust generalization across hospitals and vendors remains an open challenge, and overfitting to specific data distributions can compromise safety when deployed broadly.

Finally, deep generative models typically lack clear mechanisms for providing worst-case reconstruction guarantees. For long-term archival and interoperability, standards and clinical workflows generally favor predictable and auditable behavior; current generative approaches rarely provide bounded-error guarantees or standardized codestream definitions, which restricts their use primarily to research settings.

\subsection{Challenges in Clinical Workflow Integration}

While deep and generative models represent promising research directions, integrating learning-based compression algorithms in realistic clinical workflows still faces several major challenges:

Existing studies frequently rely on PSNR, SSIM, or MS-SSIM
as primary metrics, but these correlate imperfectly with diagnostic quality, and may
fail to capture clinically relevant distortions~\cite{cosman2002evaluating}. Clinical validation requires radiologist-in-the-loop assessment with task-specific
endpoints. For example, the detectability of low-contrast liver lesions in CT,
microbleeds in SWI, subtle wall thickening in MRI, or vessel lumen clarity in CTA may
be disproportionately degraded even when global metrics appear high. Such studies
are expensive, require IRB approval, and often adhere to strict reader-study
protocols that are not feasible for most engineering research groups.

Variability in acquisition protocols and reconstruction kernels further complicates
evaluation. A codec optimized for a Siemens CT kernel may not generalize to GE or
Philips scanners. Moreover, the presence of post-processing filters (e.g.,
denoising or beam-hardening correction) may alter redundancy characteristics,
requiring modality-specific codec tuning.

Finally, transitioning research prototypes into mature clinical solutions requires
compatibility with existing PACS, VNA, and DICOM
standards. i) DICOM Part 5 defines structured transfer syntaxes for conventional codecs such as JPEG, JPEG-LS, JPEG 2000, and RLE, but does not currently define a standard transfer syntax for neural network–based compression. DL-based codecs typically produce latent representations, weight files, or model
parameters that are not directly compatible with existing standard transfer syntaxes.
Although DICOM encapsulated objects support arbitrary binary blobs, such solutions
are not interoperable with standard viewers, preventing clinical deployment. ii) Clinical systems require backward compatibility and long-term readability.
A model trained in 2024 may not be executable in 2034 if frameworks, hardware,
or dependencies evolve. This breaks archival guarantees required for compliance
with ISO 12052 and hospital data-governance policies.

\color{black}

\section{Future Directions}

\textbf{Universal Compressor aided by Large Models}: Leveraging large foundation models (LFMs, \cite{thirunavukarasu2023large}) to create a new paradigm of universal medical image compression is compelling, promising unprecedented adaptability and efficiency across diverse medical image types. Unlike traditional compressors limited by predefined statistical models or data-specific assumptions, an LFM-aided universal compressor would utilize powerful learned priors and pattern recognition capabilities embedded within pre-trained giants like LLMs or multimodal models. By treating compression as a high-level semantic understanding and reconstruction task, such a system could dynamically infer latent structures capturing complex long-range dependencies and contextual nuances that are potentially difficult for conventional algorithms to model explicitly. Initial implementations would likely adopt a hybrid framework: the LFM generates compact, high-fidelity ``latent representations'' or predictive tokens that distill essential information, while a lightweight classical encoder (e.g., an entropy coder) efficiently packages these tokens. Challenges remain, notably the computational intensity of inference and the need for specialized fine-tuning to optimize compression ratios versus distortion. However, as model efficiency improves and techniques like knowledge distillation \cite{gou2021knowledge} advance, this approach could move closer to a ``one-size-fits-all'' compressor—simplifying data pipelines, enabling novel cross-modal medical image compression, and reducing storage and bandwidth demands for increasingly heterogeneous digital ecosystems.

\textbf{Dataset Distillation}: Dataset distillation (also called dataset condensation or dataset synthesis, \cite{lei2023comprehensive}) is a machine learning technique that synthesizes a small, highly informative synthetic dataset designed to train models nearly as effectively as the original, much larger dataset. Dataset distillation holds significant promise for revolutionizing medical AI by synthesizing ultra-compact, high-fidelity surrogate datasets that capture essential statistical and diagnostic information from large-scale imaging repositories. By distilling thousands of MRI, CT, or histopathology scans into a small cohort of synthetic ``support images'' paired with optimized labels, this approach could accelerate model training, reduce computational burdens, and democratize access to performant medical AI—particularly for resource-constrained institutions. Crucially, distilled datasets may help mitigate privacy and data-sharing hurdles inherent to sensitive patient information, thereby facilitating responsible data sharing. Ultimately, distilled medical datasets could alleviate data scarcity when training large models, transforming how the healthcare ecosystem leverages imaging data.

\textbf{Quantum Computing}: The integration of quantum computing \cite{knill2010quantum} into medical image compression holds transformative potential. Quantum approaches could exploit superposition and entanglement to process vast datasets faster than classical systems, potentially accelerating compression of multi-gigabyte MRI or CT scans. Techniques like the Quantum Fourier Transform (QFT, \cite{weinstein2001implementation}) or quantum wavelet transforms \cite{li2021multilevel} may optimize frequency-domain compression, while quantum machine learning models could intelligently prioritize diagnostically critical regions—minimizing loss in key anatomical features. Hybrid quantum-classical frameworks will likely bridge current hardware limitations, using quantum processors for selected optimization or encoding subroutines while offloading other components to classical systems. Challenges such as qubit stability, error correction, and quantum-to-classical data conversion remain, but advancements in fault-tolerant quantum hardware and quantum RAM (qRAM) could unlock ultra-efficient, high-fidelity compression.

\section{Conclusion}

This survey summarized medical image compression techniques for 2D and volumetric (3D/4D) imaging, covering both traditional and learning-based paradigms. Medical imaging imposes distinctive constraints beyond natural images, including high bit depth, low-contrast pathologies, anisotropic volumetric sampling, and strict requirements on determinism, traceability, and archival interoperability.

Classical pipelines built on transforms and predictive coding remain dominant in clinical systems because they are standardized, interpretable, and reliable, and they continue to evolve through context modeling and ROI-/anatomy-aware refinements. Learning-based codecs broaden the design space by learning compact representations and entropy models and have reported strong rate--distortion performance in many settings, while INR-based methods offer resolution-flexible decoding by representing data as continuous functions but face practical barriers in compute cost and clinical validation.

Given heterogeneous modalities and tasks, no single compressor is optimal for all medical scenarios. Hybrid designs that integrate established pipelines with learned components are a pragmatic direction toward balancing compression efficiency with clinically verifiable fidelity. Key open challenges include standardization in DICOM/PACS, efficient deployment for large-scale 3D/4D data, task-driven evaluation beyond global distortion metrics, and regulatory and ethical compliance.

\section*{Acknowledgments}
All authors declare that they have no known conflicts of interest in terms of competing financial interests or personal relationships that could have an influence or are relevant to the work reported in this paper.

\bibliographystyle{IEEEtran}
\bibliography{references}

@article{rontgen1895, author = {W. C. Röntgen}, title = {Über eine neue Art von Strahlen}, journal = {Sitzungsberichte der Physikalisch-Medizinischen Gesellschaft zu Würzburg}, pages = {132--141}, month = {Dec}, year = {1895} }

@article{seibert1995, author = {J. A. Seibert}, title = {One hundred years of medical diagnostic imaging technology}, journal = {Health Physics}, volume = {69}, number = {5}, pages = {695--720}, month = {Nov}, year = {1995}, doi = {10.1097/00004032-199511000-00006} }

@article{hounsfield1973, author = {G. N. Hounsfield}, title = {Computerized transverse axial scanning (tomography): Part 1. Description of system}, journal = {British Journal of Radiology}, volume = {46}, number = {552}, pages = {1016--1022}, month = {Dec}, year = {1973}, doi = {10.1259/0007-1285-46-552-1016} }

@article{lauterbur1973, author = {P. C. Lauterbur}, title = {Image formation by induced local interactions: Examples employing nuclear magnetic resonance}, journal = {Nature}, volume = {242}, number = {5394}, pages = {190--191}, month = {Mar}, year = {1973}, doi = {10.1038/242190a0} }

@article{me2012survey,
  title={A survey on various compression methods for medical images},
  author={Me, S Sridevi and Vijayakuymar, VR and Anuja, R},
  journal={International Journal of Intelligent Systems and Applications (IJISA)},
  volume={4},
  number={3},
  pages={13},
  year={2012}
}

@misc{dicom_standard,
  title = {Digital Imaging and Communications in Medicine (DICOM) Standard},
  author = {{National Electrical Manufacturers Association}},
  note = {NEMA PS3 / ISO 12052},
  address = {Rosslyn, VA, USA},
  howpublished = {Available free at \url{http://www.dicomstandard.org}}
}

@article{gou2021knowledge,
  title={Knowledge distillation: A survey},
  author={Gou, Jianping and Yu, Baosheng and Maybank, Stephen J and Tao, Dacheng},
  journal={International Journal of Computer Vision},
  volume={129},
  number={6},
  pages={1789--1819},
  year={2021},
  publisher={Springer}
}

@article{thirunavukarasu2023large,
  title={Large language models in medicine},
  author={Thirunavukarasu, Arun James and Ting, Darren Shu Jeng and Elangovan, Kabilan and Gutierrez, Laura and Tan, Ting Fang and Ting, Daniel Shu Wei},
  journal={Nature medicine},
  volume={29},
  number={8},
  pages={1930--1940},
  year={2023},
  publisher={Nature Publishing Group US New York}
}

@article{dupont2021coin,
  title={Coin: Compression with implicit neural representations},
  author={Dupont, Emilien and Goli{\'n}ski, Adam and Alizadeh, Milad and Teh, Yee Whye and Doucet, Arnaud},
  journal={arXiv preprint arXiv:2103.03123},
  year={2021}
}

@article{dupont2022coin++,
  title={Coin++: Neural compression across modalities},
  author={Dupont, Emilien and Loya, Hrushikesh and Alizadeh, Milad and Goli{\'n}ski, Adam and Teh, Yee Whye and Doucet, Arnaud},
  journal={arXiv preprint arXiv:2201.12904},
  year={2022}
}

@inproceedings{strumpler2022implicit,
  title={Implicit neural representations for image compression},
  author={Str{\"u}mpler, Yannick and Postels, Janis and Yang, Ren and Gool, Luc Van and Tombari, Federico},
  booktitle={European Conference on Computer Vision},
  pages={74--91},
  year={2022},
  organization={Springer}
}

@article{sanjeet2024breaking,
  title={Breaking the Barriers of One-to-One Usage of Implicit Neural Representation in Image Compression: A Linear Combination Approach with Performance Guarantees},
  author={Sanjeet, Sai and Hosseinalipour, Seyyedali and Xiong, Jinjun and Fujita, Masahiro and Sahoo, Bibhu Datta},
  journal={IEEE Internet of Things Journal},
  year={2024},
  publisher={IEEE}
}

@inproceedings{rezasoltani2023hyperspectral,
  title={Hyperspectral image compression using implicit neural representations},
  author={Rezasoltani, Shima and Qureshi, Faisal Z},
  booktitle={2023 20th Conference on Robots and Vision (CRV)},
  pages={248--255},
  year={2023},
  organization={IEEE}
}

@article{chen2021nerv,
  title={Nerv: Neural representations for videos},
  author={Chen, Hao and He, Bo and Wang, Hanyu and Ren, Yixuan and Lim, Ser Nam and Shrivastava, Abhinav},
  journal={Advances in Neural Information Processing Systems},
  volume={34},
  pages={21557--21568},
  year={2021}
}

@inproceedings{zheng2024hybrid,
  title={Hybrid Representation for 4D Medical Image Compression},
  author={Zheng, Wuyang and Meng, Jiarui and Zhang, Jiaqi and Zhang, Jian and Ma, Siwei},
  booktitle={2024 IEEE International Conference on Visual Communications and Image Processing (VCIP)},
  pages={1--5},
  year={2024},
  organization={IEEE}
}

@article{ma2024semantic,
  title={Semantic redundancy-aware implicit neural compression for multidimensional biomedical image data},
  author={Ma, Yifan and Yi, Chengqiang and Zhou, Yao and Wang, Zhaofei and Zhao, Yuxuan and Zhu, Lanxin and Wang, Jie and Gao, Shimeng and Liu, Jianchao and Yuan, Xinyue and others},
  journal={Communications Biology},
  volume={7},
  number={1},
  pages={1081},
  year={2024},
  publisher={Nature Publishing Group UK London}
}

@inproceedings{yang2023sci,
  title={Sci: A spectrum concentrated implicit neural compression for biomedical data},
  author={Yang, Runzhao and Xiao, Tingxiong and Cheng, Yuxiao and Cao, Qianni and Qu, Jinyuan and Suo, Jinli and Dai, Qionghai},
  booktitle={Proceedings of the AAAI Conference on Artificial Intelligence},
  volume={37},
  number={4},
  pages={4774--4782},
  year={2023}
}

@article{dai2024implicit,
  title={Implicit Neural Image Field for Biological Microscopy Image Compression},
  author={Dai, Gaole and Tseng, Cheng-Ching and Wuwu, Qingpo and Zhang, Rongyu and Wang, Shaokang and Lu, Ming and Huang, Tiejun and Zhou, Yu and Tuz, Ali Ata and Gunzer, Matthias and others},
  journal={arXiv preprint arXiv:2405.19012},
  year={2024}
}

@inproceedings{yang2023tinc2,
  title={Tinc: Tree-structured implicit neural compression},
  author={Yang, Runzhao},
  booktitle={Proceedings of the IEEE/CVF Conference on Computer Vision and Pattern Recognition},
  pages={18517--18526},
  year={2023}
}

@article{sitzmann2020implicit,
  title={Implicit neural representations with periodic activation functions},
  author={Sitzmann, Vincent and Martel, Julien and Bergman, Alexander and Lindell, David and Wetzstein, Gordon},
  journal={Advances in neural information processing systems},
  volume={33},
  pages={7462--7473},
  year={2020}
}

@article{mildenhall2021nerf,
  title={Nerf: Representing scenes as neural radiance fields for view synthesis},
  author={Mildenhall, Ben and Srinivasan, Pratul P and Tancik, Matthew and Barron, Jonathan T and Ramamoorthi, Ravi and Ng, Ren},
  journal={Communications of the ACM},
  volume={65},
  number={1},
  pages={99--106},
  year={2021},
  publisher={ACM New York, NY, USA}
}

@article{li2025towards,
  title={Towards scalable medical image compression using hybrid model analysis},
  author={Li, Shunlei and Lu, Jiajie and Hu, Yingbai and Mattos, Leonardo S and Li, Zheng},
  journal={Journal of Big Data},
  volume={12},
  number={1},
  pages={45},
  year={2025},
  publisher={Springer}
}

@article{sheibanifard2023novel,
  title={A novel implicit neural representation for volume data},
  author={Sheibanifard, Armin and Yu, Hongchuan},
  journal={Applied Sciences},
  volume={13},
  number={5},
  pages={3242},
  year={2023},
  publisher={MDPI}
}

@inproceedings{gao2023sinco,
  title={Sinco: A novel structural regularizer for image compression using implicit neural representations},
  author={Gao, Harry and Gan, Weijie and Sun, Zhixin and Kamilov, Ulugbek S},
  booktitle={ICASSP 2023-2023 IEEE International Conference on Acoustics, Speech and Signal Processing (ICASSP)},
  pages={1--5},
  year={2023},
  organization={IEEE}
}

@article{yang2024unicompress,
  title={Unicompress: Enhancing multi-data medical image compression with knowledge distillation},
  author={Yang, Runzhao and Chen, Yinda and Zhang, Zhihong and Liu, Xiaoyu and Li, Zongren and He, Kunlun and Xiong, Zhiwei and Suo, Jinli and Dai, Qionghai},
  journal={arXiv preprint arXiv:2405.16850},
  year={2024}
}

@article{min2022lossless,
  title={Lossless medical image compression based on anatomical information and deep neural networks},
  author={Min, Qiusha and Wang, Xin and Huang, Bo and Zhou, Zhongwei},
  journal={Biomedical Signal Processing and Control},
  volume={74},
  pages={103499},
  year={2022},
  publisher={Elsevier}
}

@article{luu2021efficiently,
  title={Efficiently compressing 3D medical images for teleinterventions via CNNs and anisotropic diffusion},
  author={Luu, Ha Manh and van Walsum, Theo and Franklin, Daniel and Pham, Phuong Cam and Vu, Luu Dang and Moelker, Adriaan and Staring, Marius and VanHoang, Xiem and Niessen, Wiro and Trung, Nguyen Linh},
  journal={Medical Physics},
  volume={48},
  number={6},
  pages={2877--2890},
  year={2021},
  publisher={Wiley Online Library}
}

@article{chen2023streaming,
  title={Streaming Lossless Volumetric Compression of Medical Images Using Gated Recurrent Convolutional Neural Network},
  author={Chen, Qianhao and Chen, Jietao},
  journal={arXiv preprint arXiv:2311.16200},
  year={2023}
}

@inproceedings{nagoor2021medzip,
  title={MedZip: 3D medical images lossless compressor using recurrent neural network (LSTM)},
  author={Nagoor, Omniah H and Whittle, Joss and Deng, Jingjing and Mora, Benjamin and Jones, Mark W},
  booktitle={2020 25th international conference on pattern recognition (ICPR)},
  pages={2874--2881},
  year={2021},
  organization={IEEE}
}

@article{nagoor2022sampling,
  title={Sampling strategies for learning-based 3D medical image compression},
  author={Nagoor, Omniah H and Whittle, Joss and Deng, Jingjing and Mora, Benjamin and Jones, Mark W},
  journal={Machine Learning with Applications},
  volume={8},
  pages={100273},
  year={2022},
  publisher={Elsevier}
}

@article{wang2024learning,
  title={Learning Lossless Compression for High Bit-Depth Volumetric Medical Image},
  author={Wang, Kai and Bai, Yuanchao and Li, Daxin and Zhai, Deming and Jiang, Junjun and Liu, Xianming},
  journal={IEEE Transactions on Image Processing},
  year={2024},
  publisher={IEEE}
}

@article{bruylants2009jp3d,
  title={JP3D--extensions for three-dimensional data (part 10)},
  author={Bruylants, Tim and Schelkens, Peter and Tzannes, Alexis},
  journal={The JPEG 2000 Suite},
  pages={199--227},
  year={2009},
  publisher={Wiley-Blackwell}
}

@article{vspelivc2012lossless,
  title={Lossless compression of threshold-segmented medical images},
  author={{\v{S}}peli{\v{c}}, Denis and {\v{Z}}alik, Borut},
  journal={Journal of medical systems},
  volume={36},
  pages={2349--2357},
  year={2012},
  publisher={Springer}
}

@article{sanchez2009symmetry,
  title={Symmetry-based scalable lossless compression of 3D medical image data},
  author={Sanchez, Victor and Abugharbieh, Rafeef and Nasiopoulos, Panos},
  journal={IEEE Transactions on Medical Imaging},
  volume={28},
  number={7},
  pages={1062--1072},
  year={2009},
  publisher={IEEE}
}

@article{parikh2017high,
  title={High bit-depth medical image compression with HEVC},
  author={Parikh, Saurin S and Ruiz, Damian and Kalva, Hari and Fern{\'a}ndez-Escribano, Gerardo and Adzic, Velibor},
  journal={IEEE journal of biomedical and health informatics},
  volume={22},
  number={2},
  pages={552--560},
  year={2017},
  publisher={IEEE}
}

@article{liu2024bilateral,
  title={Bilateral context modeling for residual coding in lossless 3D medical image compression},
  author={Liu, Xiangrui and Wang, Meng and Wang, Shiqi and Kwong, Sam},
  journal={IEEE Transactions on Image Processing},
  year={2024},
  publisher={IEEE}
}

@article{xue2022aiwave,
  title={AiWave: Volumetric image compression with 3-D trained affine wavelet-like transform},
  author={Xue, Dongmei and Ma, Haichuan and Li, Li and Liu, Dong and Xiong, Zhiwei},
  journal={IEEE Transactions on Medical Imaging},
  volume={42},
  number={3},
  pages={606--618},
  year={2022},
  publisher={IEEE}
}

@article{ahmed2006discrete,
  title={Discrete cosine transform},
  author={Ahmed, Nasir and Natarajan, T\_ and Rao, Kamisetty R},
  journal={IEEE transactions on Computers},
  volume={100},
  number={1},
  pages={90--93},
  year={2006},
  publisher={IEEE}
}

@article{chen2022exploiting,
  title={Exploiting intra-slice and inter-slice redundancy for learning-based lossless volumetric image compression},
  author={Chen, Zhenghao and Gu, Shuhang and Lu, Guo and Xu, Dong},
  journal={IEEE Transactions on Image Processing},
  volume={31},
  pages={1697--1707},
  year={2022},
  publisher={IEEE}
}

@ARTICLE{DCT1974,
  author={Ahmed, N. and Natarajan, T. and Rao, K.R.},
  journal={IEEE Transactions on Computers}, 
  title={Discrete Cosine Transform}, 
  year={1974},
  volume={C-23},
  number={1},
  pages={90-93},
  doi={10.1109/T-C.1974.223784}}

@article{taubman2002jpeg2000,
  title={JPEG2000: Standard for interactive imaging},
  author={Taubman, David S and Marcellin, Michael W},
  journal={Proceedings of the IEEE},
  volume={90},
  number={8},
  pages={1336--1357},
  year={2002},
  publisher={IEEE}
}

@inproceedings{muraki1992approximation,
  title={Approximation and rendering of volume data using wavelet transforms},
  author={Muraki, Shigeru},
  booktitle={Proceedings Visualization'92},
  pages={21--22},
  year={1992},
  organization={IEEE Computer Society}
}

@inproceedings{kim1999lossless,
  title={Lossless volumetric medical image compression},
  author={Kim, Young-Seop and Pearlman, William A},
  booktitle={Applications of Digital Image Processing XXII},
  volume={3808},
  pages={305--312},
  year={1999},
  organization={SPIE}
}

@article{senapati2016volumetric,
  title={Volumetric medical image compression using 3D listless embedded block partitioning},
  author={Senapati, Ranjan K and Prasad, PM K and Swain, Gandharba and Shankar, TN},
  journal={SpringerPlus},
  volume={5},
  number={1},
  pages={2100},
  year={2016},
  publisher={Springer}
}

@inproceedings{bernabe2000new,
  title={A new lossy 3-D wavelet transform for high-quality compression of medical video},
  author={Bernab{\'e}, Gregorio and Gonz{\'a}lez, Jos{\'e} and Garcia, Jose Manuel and Duato, Jose},
  booktitle={Proceedings 2000 IEEE EMBS International Conference on Information Technology Applications in Biomedicine. ITAB-ITIS 2000. Joint Meeting Third IEEE EMBS International Conference on Information Technol},
  pages={226--231},
  year={2000},
  organization={IEEE}
}

@article{angelidis1994mr,
  title={MR image compression using a wavelet transform coding algorithm},
  author={Angelidis, PA},
  journal={Magnetic Resonance Imaging},
  volume={12},
  number={7},
  pages={1111--1120},
  year={1994},
  publisher={Elsevier}
}

@article{ammah2019robust,
  title={Robust medical image compression based on wavelet transform and vector quantization},
  author={Ammah, Paul Nii Tackie and Owusu, Ebenezer},
  journal={Informatics in medicine unlocked},
  volume={15},
  pages={100183},
  year={2019},
  publisher={Elsevier}
}

@article{linde2003algorithm,
  title={An algorithm for vector quantizer design},
  author={Linde, Yoseph and Buzo, Andres and Gray, Robert},
  journal={IEEE Transactions on communications},
  volume={28},
  number={1},
  pages={84--95},
  year={2003},
  publisher={IEEE}
}

@article{jiang2012medical,
  title={Medical image compression based on vector quantization with variable block sizes in wavelet domain},
  author={Jiang, Huiyan and Ma, Zhiyuan and Hu, Yang and Yang, Benqiang and Zhang, Libo},
  journal={Computational intelligence and neuroscience},
  volume={2012},
  number={1},
  pages={541890},
  year={2012},
  publisher={Wiley Online Library}
}

@article{bruylants2015wavelet,
  title={Wavelet based volumetric medical image compression},
  author={Bruylants, Tim and Munteanu, Adrian and Schelkens, Peter},
  journal={Signal processing: Image communication},
  volume={31},
  pages={112--133},
  year={2015},
  publisher={Elsevier}
}

@inproceedings{raj2007novel,
  title={A novel approach to medical image compression using sequential 3D DCT},
  author={Raj, V Naga Prudhvi and Venkateswarlu, T},
  booktitle={International Conference on Computational Intelligence and Multimedia Applications (ICCIMA 2007)},
  volume={3},
  pages={146--152},
  year={2007},
  organization={IEEE}
}

@article{xue20213d,
  title={3D DCT based image compression method for the medical endoscopic application},
  author={Xue, Jiawen and Yin, Li and Lan, Zehua and Long, Mingzhu and Li, Guolin and Wang, Zhihua and Xie, Xiang},
  journal={Sensors},
  volume={21},
  number={5},
  pages={1817},
  year={2021},
  publisher={MDPI}
}

@inproceedings{sanchez2008efficient,
  title={Efficient 4D motion compensated lossless compression of dynamic volumetric medical image data},
  author={Sanchez, Victor and Nasiopoulos, Panos and Abugharbieh, Rafeef},
  booktitle={2008 IEEE International Conference on Acoustics, Speech and Signal Processing},
  pages={549--552},
  year={2008},
  organization={IEEE}
}

@inproceedings{sanchez2006lossless,
  title={Lossless compression of 4D medical images using H. 264/AVC},
  author={Sanchez, Victor and Nasiopoulos, Panos and Abugharbieh, Rafeef},
  booktitle={2006 IEEE International Conference on Acoustics Speech and Signal Processing Proceedings},
  volume={2},
  pages={II--II},
  year={2006},
  organization={IEEE}
}

@inproceedings{martin2008analysis,
  title={Analysis of compression of 4D volumetric medical image datasets using multi-view (MVC) video coding methods},
  author={Martin, Uwe-Erik and Kaup, Andr{\'e}},
  booktitle={Mathematics of Data/Image Pattern Recognition, Compression, and Encryption with Applications XI},
  volume={7075},
  pages={56--63},
  year={2008},
  organization={SPIE}
}

@article{kassim2005motion,
  title={Motion compensated lossy-to-lossless compression of 4-D medical images using integer wavelet transforms},
  author={Kassim, Ashraf A and Yan, Pingkun and Lee, Wei Siong and Sengupta, Kuntal},
  journal={IEEE transactions on information technology in biomedicine},
  volume={9},
  number={1},
  pages={132--138},
  year={2005},
  publisher={IEEE}
}

@article{lucas2017lossless,
  title={Lossless compression of medical images using 3-D predictors},
  author={Lucas, Lu{\'\i}s FR and Rodrigues, Nuno MM and da Silva Cruz, Luis A and de Faria, S{\'e}rgio MM},
  journal={IEEE transactions on medical imaging},
  volume={36},
  number={11},
  pages={2250--2260},
  year={2017},
  publisher={IEEE}
}

@inproceedings{santos2015contributions,
  title={Contributions to lossless coding of medical images using minimum rate predictors},
  author={Santos, Jo{\~a}o M and Guarda, Andr{\'e} FR and Rodrigues, Nuno MM and Faria, S{\'e}rgio MM},
  booktitle={2015 IEEE International Conference on Image Processing (ICIP)},
  pages={2935--2939},
  year={2015},
  organization={IEEE}
}

@inproceedings{santos2016compression,
  title={Compression of medical images using MRP with bi-directional prediction and histogram packing},
  author={Santos, Joao M and Guarda, Andr{\'e} FR and da Silva Cruz, Lu{\'\i}s A and Rodrigues, Nuno MM and Faria, S{\'e}rgio MM},
  booktitle={2016 Picture Coding Symposium (PCS)},
  pages={1--5},
  year={2016},
  organization={IEEE}
}

@inproceedings{diez2005lossless,
  title={A lossless compression algorithm based on predictive coding for volumetric medical datasets},
  author={D{\'\i}ez-Garc{\'\i}a, M{\'o}nica and Simmross-Wattenberg, Federico and Alberola-L{\'o}pez, Carlos},
  booktitle={2005 13th European Signal Processing Conference},
  pages={1--4},
  year={2005},
  organization={IEEE}
}

@article{guarda2017method,
  title={A method to improve HEVC lossless coding of volumetric medical images},
  author={Guarda, Andr{\'e} FR and Santos, Jo{\~a}o M and da Silva Cruz, Lu{\'\i}s A and Assun{\c{c}}{\~a}o, Pedro AA and Rodrigues, Nuno MM and de Faria, S{\'e}rgio MM},
  journal={Signal Processing: Image Communication},
  volume={59},
  pages={96--104},
  year={2017},
  publisher={Elsevier}
}

@article{song2018lossless,
  title={Lossless medical image compression using geometry-adaptive partitioning and least square-based prediction},
  author={Song, Xiaoying and Huang, Qijun and Chang, Sheng and He, Jin and Wang, Hao},
  journal={Medical \& biological engineering \& computing},
  volume={56},
  pages={957--966},
  year={2018},
  publisher={Springer}
}

@article{mallat2002theory,
  title={A theory for multiresolution signal decomposition: the wavelet representation},
  author={Mallat, Stephane G},
  journal={IEEE transactions on pattern analysis and machine intelligence},
  volume={11},
  number={7},
  pages={674--693},
  year={2002},
  publisher={Ieee}
}

@book{suetens2017fundamentals,
  title={Fundamentals of medical imaging},
  author={Suetens, Paul},
  year={2017},
  publisher={Cambridge university press}
}

@book{cover1999elements,
  title={Elements of information theory},
  author={Cover, Thomas M},
  year={1999},
  publisher={John Wiley \& Sons}
}

@article{devore2002image,
  title={Image compression through wavelet transform coding},
  author={DeVore, Ronald A and Jawerth, Bj{\"o}rn and Lucier, Bradley J},
  journal={IEEE Transactions on information theory},
  volume={38},
  number={2},
  pages={719--746},
  year={2002},
  publisher={IEEE}
}

@article{lewis1992image,
  title={Image compression using the 2-D wavelet transform},
  author={Lewis, Adrian S and Knowles, G},
  journal={IEEE Transactions on image Processing},
  volume={1},
  number={2},
  pages={244--250},
  year={1992},
  publisher={IEEE}
}

@article{shapiro2002embeddedsaid1996new,
  title={Embedded image coding using zerotrees of wavelet coefficients},
  author={Shapiro, Jerome M},
  journal={IEEE Transactions on signal processing},
  volume={41},
  number={12},
  pages={3445--3462},
  year={2002},
  publisher={IEEE}
}

@article{farghaly2020floating,
  title={Floating-point discrete wavelet transform-based image compression on FPGA},
  author={Farghaly, Sarah H and Ismail, Samar M},
  journal={AEU-International Journal of Electronics and Communications},
  volume={124},
  pages={153363},
  year={2020},
  publisher={Elsevier}
}

@article{paul2022health,
  title={A health care image compression scheme using discrete wavelet transform and convolution neural network},
  author={Paul, Raj Kumar and Chandran, Saravanan},
  journal={J Eng Res. https://doi. org/10.36909/jer. ICMET},
  volume={17163},
  year={2022}
}

@article{viswanathan2024empirical,
  title={An Empirical Selection of Wavelet for Near-lossless Medical Image Compression},
  author={Viswanathan, Punitha and Palanisamy, Kalavathi},
  journal={Current Medical Imaging},
  volume={20},
  pages={e300323215226},
  year={2024},
  publisher={Bentham Science Publishers},
  doi={10.2174/1573405620666230330113833}
}

@article{wallace1991jpeg,
  title={The JPEG still picture compression standard},
  author={Wallace, Gregory K},
  journal={Communications of the ACM},
  volume={34},
  number={4},
  pages={30--44},
  year={1991},
  publisher={AcM New York, NY, USA}
}

@article{singh2007dwt,
  title={DWT--DCT hybrid scheme for medical image compression},
  author={Singh, S and Kumar, V and Verma, HK},
  journal={Journal of medical engineering \& technology},
  volume={31},
  number={2},
  pages={109--122},
  year={2007},
  publisher={Taylor \& Francis}
}

@article{chen2007medical,
  title={Medical image compression using DCT-based subband decomposition and modified SPIHT data organization},
  author={Chen, Yen-Yu},
  journal={International journal of medical informatics},
  volume={76},
  number={10},
  pages={717--725},
  year={2007},
  publisher={Elsevier}
}

@inproceedings{chikouche2008application,
  title={Application of the DCT and arithmetic coding to medical image compression},
  author={Chikouche, Djamel and Benzid, Ridha and Bentoumi, Miloud},
  booktitle={2008 3rd International Conference on Information and Communication Technologies: From Theory to Applications},
  pages={1--5},
  year={2008},
  organization={IEEE}
}

@article{yadav2023near,
  title={Near-lossless compression of Tc-99 m DMSA scan images using discrete cosine transformation},
  author={Yadav, Priya and Pandey, Anil Kumar and Chaudhary, Jagrati and Sharma, Param Dev and Jaleel, Jasim and Baghel, Vivek and Patel, Chetan and Kumar, Rakesh},
  journal={Indian Journal of Nuclear Medicine},
  volume={38},
  number={3},
  pages={231--238},
  year={2023},
  publisher={Medknow}
}

@article{ting2015novel,
  title={A novel approach for arbitrary-shape roi compression of medical images using principal component analysis (pca)},
  author={Ting, Lim Sin and Weng, David Yap Fook and Manap, Nurulfajar Bin Abdul},
  journal={Trends in Applied Sciences Research},
  volume={10},
  number={1},
  pages={68},
  year={2015},
  publisher={Academic Journals Inc.}
}

@article{reddy2018new,
  title={A new approach for the image compression to the medical images using PCA-SPIHT.},
  author={Reddy, Rajasekhar M and Ravichandran, Kattur Soundarapandian and Venkatraman, B and Suganya, SD},
  journal={Biomedical Research (0970-938X)},
  year={2018}
}

@article{anandan2016medical,
  title={Medical image compression using wrapping based fast discrete curvelet transform and arithmetic coding},
  author={Anandan, P and Sabeenian, RS},
  journal={Circuits and Systems},
  volume={7},
  number={8},
  pages={2059--2069},
  year={2016},
  publisher={Scientific Research Publishing}
}

@article{hashemi2013new,
  title={A new contourlet-based compression and speckle reduction method for medical ultrasound images},
  author={Hashemi-berenjabad, Seyyed Hadi and Mahloojifar, Ali},
  journal={International Journal of Computer Applications},
  volume={82},
  number={13},
  year={2013},
  publisher={Citeseer}
}

@article{juliet2015projection,
  title={Projection-based medical image compression for telemedicine applications},
  author={Juliet, Sujitha and Rajsingh, Elijah Blessing and Ezra, Kirubakaran},
  journal={Journal of digital imaging},
  volume={28},
  pages={146--159},
  year={2015},
  publisher={Springer}
}

@article{liu2017current,
  title={The current role of image compression standards in medical imaging},
  author={Liu, Feng and Hernandez-Cabronero, Miguel and Sanchez, Victor and Marcellin, Michael W and Bilgin, Ali},
  journal={Information},
  volume={8},
  number={4},
  pages={131},
  year={2017},
  publisher={MDPI}
}

@article{rojas2022lossless,
  title={Lossless medical image compression by using difference transform},
  author={Rojas-Hern{\'a}ndez, Rafael and D{\'\i}az-de-Le{\'o}n-Santiago, Juan Luis and Barcel{\'o}-Alonso, Grettel and Bautista-L{\'o}pez, Jorge and Trujillo-Mora, Valentin and Salgado-Ram{\'\i}rez, Julio C{\'e}sar},
  journal={Entropy},
  volume={24},
  number={7},
  pages={951},
  year={2022},
  publisher={MDPI}
}

@inproceedings{weinberger1996loco,
  title={LOCO-I: A low complexity, context-based, lossless image compression algorithm},
  author={Weinberger, Marcelo J and Seroussi, Gadiel and Sapiro, Guillermo},
  booktitle={Proceedings of Data Compression Conference-DCC'96},
  pages={140--149},
  year={1996},
  organization={IEEE}
}

@article{weinberger2000loco,
  title={The LOCO-I lossless image compression algorithm: Principles and standardization into JPEG-LS},
  author={Weinberger, Marcelo J and Seroussi, Gadiel and Sapiro, Guillermo},
  journal={IEEE Transactions on Image processing},
  volume={9},
  number={8},
  pages={1309--1324},
  year={2000},
  publisher={IEEE}
}

@inproceedings{wu1996calic,
  title={CALIC-a context based adaptive lossless image codec},
  author={Wu, Xiaolin and Memon, Nasir},
  booktitle={1996 IEEE international conference on acoustics, speech, and signal processing conference proceedings},
  volume={4},
  pages={1890--1893},
  year={1996},
  organization={IEEE}
}

@article{shirsat2013lossless,
  title={Lossless medical image compression by integer wavelet and predictive coding},
  author={Shirsat, TG and Bairagi, VK},
  journal={International Scholarly Research Notices},
  volume={2013},
  number={1},
  pages={832527},
  year={2013},
  publisher={Wiley Online Library}
}

@article{puthooran2013lossless,
  title={Lossless compression of medical images using a dual level DPCM with context adaptive switching neural network predictor},
  author={Puthooran, Emjee and Anand, Radhey Shyam and Mukherjee, Shaktidev},
  journal={International Journal of Computational Intelligence Systems},
  volume={6},
  number={6},
  pages={1082--1093},
  year={2013},
  publisher={Springer}
}

@article{miaou2009lossless,
  title={A lossless compression method for medical image sequences using JPEG-LS and interframe coding},
  author={Miaou, Shaou-Gang and Ke, Fu-Sheng and Chen, Shu-Ching},
  journal={IEEE transactions on information technology in biomedicine},
  volume={13},
  number={5},
  pages={818--821},
  year={2009},
  publisher={IEEE}
}

@article{zuo2015improved,
  title={An improved medical image compression technique with lossless region of interest},
  author={Zuo, Zhiyong and Lan, Xia and Deng, Lihua and Yao, Shoukui and Wang, Xiaoping},
  journal={Optik},
  volume={126},
  number={21},
  pages={2825--2831},
  year={2015},
  publisher={Elsevier}
}

@inproceedings{eben2016region,
  title={Region-based prediction and quality measurements for medical image compression},
  author={Eben Sophia, P and Anitha, J},
  booktitle={Proceedings of Fifth International Conference on Soft Computing for Problem Solving: SocProS 2015, Volume 1},
  pages={355--366},
  year={2016},
  organization={Springer}
}

@article{baware2016medical,
  title={Medical image compression using adaptive prediction and block based entropy coding},
  author={Baware, Ekta Ashok and Save, Jagruti},
  journal={International Journal of Computer Applications},
  volume={153},
  number={9},
  pages={28--33},
  year={2016},
  publisher={Foundation of Computer Science}
}

@inproceedings{anitha2018optimized,
  title={An optimized predictive coding algorithm for medical image compression},
  author={Anitha, J and Eben Sophia, P and Jude Hemanth, D},
  booktitle={International Conference of the Sri Lanka Association for Artificial Intelligence},
  pages={315--324},
  year={2018},
  organization={Springer}
}

@article{xin2021lossless,
  title={A lossless compression method for multi-component medical images based on big data mining},
  author={Xin, Gangtao and Fan, Pingyi},
  journal={Scientific Reports},
  volume={11},
  number={1},
  pages={12372},
  year={2021},
  publisher={Nature Publishing Group UK London}
}

@article{sridhar2022optimal,
  title={Optimal medical image size reduction model creation using recurrent neural network and GenPSOWVQ},
  author={Sridhar, Chethana and Pareek, Piyush Kumar and Kalidoss, R and Jamal, Sajjad Shaukat and Shukla, Prashant Kumar and Nuagah, Stephen Jeswinde},
  journal={Journal of Healthcare Engineering},
  volume={2022},
  number={1},
  pages={2354866},
  year={2022},
  publisher={Wiley Online Library}
}

@inproceedings{cziho1998medical,
  title={Medical image compression using region-of-interest vector quantization},
  author={Czih{\'o}, Andr{\'a}s and Cazuguel, Guy and Solaiman, Basel and Roux, Christian},
  booktitle={Proceedings of the 20th Annual International Conference of the IEEE Engineering in Medicine and Biology Society. Vol. 20 Biomedical Engineering Towards the Year 2000 and Beyond (Cat. No. 98CH36286)},
  volume={3},
  pages={1277--1280},
  year={1998},
  organization={IEEE}
}

@article{mitra1998wavelet,
  title={Wavelet-based vector quantization for high-fidelity compression and fast transmission of medical images},
  author={Mitra, Sunanda and Yang, Shuyu and Kustov, Vadim},
  journal={Journal of digital imaging},
  volume={11},
  number={Suppl 2},
  pages={24--30},
  year={1998},
  publisher={Springer}
}

@article{kumar2018compression,
  title={Compression of CT images using contextual vector quantization with simulated annealing for telemedicine application},
  author={Kumar, Subbiahpillai Neelakantapillai and Lenin Fred, A and Sebastin Varghese, P},
  journal={Journal of Medical Systems},
  volume={42},
  number={11},
  pages={218},
  year={2018},
  publisher={Springer}
}

@article{gurjar2014medical,
  title={Medical Image Compression Using Wavelets And Vector Quantization For Telemedicine Application},
  author={Gurjar, AA and Korde, Neha S},
  journal={IOSR Journal of Electronics and Communication Engineering (IOSR-JECE)},
  volume={9},
  number={2},
  pages={76--81},
  year={2014}
}

@article{liu2022medical,
  title={Medical image compression based on variational autoencoder},
  author={Liu, Xuan and Zhang, Lu and Guo, Zihao and Han, Tailin and Ju, Mingchi and Xu, Bo and Liu, Hong},
  journal={Mathematical Problems in Engineering},
  volume={2022},
  number={1},
  pages={7088137},
  year={2022},
  publisher={Wiley Online Library}
}

@article{fettah2024convolutional,
  title={Convolutional Autoencoder-Based medical image compression using a novel annotated medical X-ray imaging dataset},
  author={Fettah, Amina and Menassel, Rafik and Gattal, Abdeljalil and Gattal, Abdelhak},
  journal={Biomedical Signal Processing and Control},
  volume={94},
  pages={106238},
  year={2024},
  publisher={Elsevier}
}

@inproceedings{molaei2023implicit,
  title={Implicit neural representation in medical imaging: A comparative survey},
  author={Molaei, Amirali and Aminimehr, Amirhossein and Tavakoli, Armin and Kazerouni, Amirhossein and Azad, Bobby and Azad, Reza and Merhof, Dorit},
  booktitle={Proceedings of the IEEE/CVF International Conference on Computer Vision},
  pages={2381--2391},
  year={2023}
}

@article{said1996, author = {A. Said and W. A. Pearlman}, title = {A new, fast, and efficient image codec based on set partitioning in hierarchical trees}, journal = {IEEE Transactions on Circuits and Systems for Video Technology}, volume = {6}, number = {3}, pages = {243--250}, month = {Jun}, year = {1996}, doi = {10.1109/76.499834} }

@article{kalaiselvi2021significant, author = {T. Kalaiselvi and P. Sriramakrishnan and K. Somasundaram}, title = {Significant medical image compression techniques: a review}, journal = {TELKOMNIKA}, volume = {19}, number = {5}, pages = {1612--1621}, month = {Oct}, year = {2021} }

@article{aziz2020comprehensive, author = {S. A. Aziz and S. M. Sam and N. Mohamed and others}, title = {The comprehensive review of neural network: an intelligent medical image compression for data sharing}, journal = {International Journal of Integrative Engineering}, volume = {12}, number = {7}, pages = {81--89}, year = {2020} }

@article{fettah2024deep, author = {A. Fettah and M. Koubaa and A. Ksibi and A. Ayadi and M. Jallouli}, title = {Deep learning-assisted medical image compression challenges and opportunities: systematic review}, journal = {Neural Computing and Applications}, volume = {36}, pages = {4769--4795}, year = {2024} }

@article{paul2022, author = {S. Paul and S. Chandran}, title = {A healthcare image compression scheme combining discrete wavelet transform and neural networks}, journal = {Biomedical Signal Processing and Control}, volume = {73}, pages = {103434}, year = {2022} }

@misc{dicom2021part5, title = {Digital Imaging and Communications in Medicine (DICOM) Part 5: Data Structures and Encoding}, author = {{NEMA}}, howpublished = {NEMA PS3.5-2021c}, year = {2021} }

@article{pinho2017current, author = {O. Pinho and E. Faria and M. Carvalho and A. J. Madureira}, title = {The Current Role of Image Compression Standards in Medical Imaging}, journal = {Information}, volume = {8}, number = {4}, pages = {131}, month = {Oct}, year = {2017} }

@article{ma2019image,
  title={Image and video compression with neural networks: A review},
  author={Ma, Siwei and Zhang, Xinfeng and Jia, Chuanmin and Zhao, Zhenghui and Wang, Shiqi and Wang, Shanshe},
  journal={IEEE Transactions on Circuits and Systems for Video Technology},
  volume={30},
  number={6},
  pages={1683--1698},
  year={2019},
  publisher={IEEE}
}

@article{moorthy2021deep, author = {A. K. Moorthy and A. C. Bovik}, title = {Deep Architectures for Image Compression: A Critical Review}, journal = {IEEE Signal Processing Magazine}, volume = {38}, number = {4}, pages = {17--30}, month = {Jul}, year = {2021} }

@article{liu2022learning, author = {D. Liu and H. Li and K. Wang and L. Li and others}, title = {Learning-driven lossy image compression: A comprehensive survey}, journal = {IEEE Transactions on Image Processing}, volume = {31}, pages = {478--492}, year = {2022} }

@article{mishra2022deep,
  title={Deep architectures for image compression: a critical review},
  author={Mishra, Dipti and Singh, Satish Kumar and Singh, Rajat Kumar},
  journal={Signal Processing},
  volume={191},
  pages={108346},
  year={2022},
  publisher={Elsevier}
}

@article{boopathiraja2022computational,
  title={Computational 2D and 3D medical image data compression models},
  author={Boopathiraja, S and Punitha, V and Kalavathi, P and Prasath, VB Surya},
  journal={Archives of Computational Methods in Engineering},
  volume={29},
  number={2},
  pages={975--1007},
  year={2022},
  publisher={Springer}
}

@article{bourai2024deep,
  title={Deep learning-assisted medical image compression challenges and opportunities: systematic review},
  author={Bourai, Nour El Houda and Merouani, Hayet Farida and Djebbar, Akila},
  journal={Neural Computing and Applications},
  volume={36},
  number={17},
  pages={10067--10108},
  year={2024},
  publisher={Springer}
}

@article{xiong2003lossy,
  title={Lossy-to-lossless compression of medical volumetric data using three-dimensional integer wavelet transforms},
  author={Xiong, Zixiang and Wu, Xiaolin and Cheng, Samuel and Hua, Jianping},
  journal={IEEE Transactions on Medical Imaging},
  volume={22},
  number={3},
  pages={459--470},
  year={2003},
  publisher={IEEE}
}

@book{huang2011pacs,
  title={PACS and imaging informatics: basic principles and applications},
  author={Huang, Hai K},
  year={2011},
  publisher={John Wiley \& Sons}
}

@article{topol2019high,
  title={High-performance medicine: the convergence of human and artificial intelligence},
  author={Topol, Eric J},
  journal={Nature medicine},
  volume={25},
  number={1},
  pages={44--56},
  year={2019},
  publisher={Nature Publishing Group US New York}
}

@article{li2021multilevel,
  title={Multilevel 2-d quantum wavelet transforms},
  author={Li, Hai-Sheng and Fan, Ping and Peng, Huiling and Song, Shuxiang and Long, Gui-Lu},
  journal={IEEE Transactions on Cybernetics},
  volume={52},
  number={8},
  pages={8467--8480},
  year={2021},
  publisher={IEEE}
}

@article{weinstein2001implementation,
  title={Implementation of the quantum Fourier transform},
  author={Weinstein, Yaakov S and Pravia, MA and Fortunato, EM and Lloyd, Seth and Cory, David G},
  journal={Physical review letters},
  volume={86},
  number={9},
  pages={1889},
  year={2001},
  publisher={APS}
}

@article{knill2010quantum,
  title={Quantum computing},
  author={Knill, Emanuel},
  journal={Nature},
  volume={463},
  number={7280},
  pages={441--443},
  year={2010},
  publisher={Nature Publishing Group UK London}
}

@article{lei2023comprehensive,
  title={A comprehensive survey of dataset distillation},
  author={Lei, Shiye and Tao, Dacheng},
  journal={IEEE Transactions on Pattern Analysis and Machine Intelligence},
  volume={46},
  number={1},
  pages={17--32},
  year={2023},
  publisher={IEEE}
}

@article{cosman2002evaluating,
  title={Evaluating quality of compressed medical images: SNR, subjective rating, and diagnostic accuracy},
  author={Cosman, Pamela C and Gray, Robert M and Olshen, Richard A},
  journal={Proceedings of the IEEE},
  volume={82},
  number={6},
  pages={919--932},
  year={2002},
  publisher={IEEE}
}

@article{langer2011challenges,
  title={Challenges for data storage in medical imaging research},
  author={Langer, Steve G},
  journal={Journal of digital imaging},
  volume={24},
  pages={203--207},
  year={2011},
  publisher={Springer}
}

@misc{wang2018improving,
  title={Improving spatial resolution at CT: development, benefits, and pitfalls},
  author={Wang, Jia and Fleischmann, Dominik},
  journal={Radiology},
  volume={289},
  number={1},
  pages={261--262},
  year={2018},
  publisher={Radiological Society of North America}
}

@article{slavin2005spatial,
  title={Spatial and temporal resolution in cardiovascular MR imaging: review and recommendations},
  author={Slavin, Glenn S and Bluemke, David A},
  journal={Radiology},
  volume={234},
  number={2},
  pages={330--338},
  year={2005},
  publisher={Radiological Society of North America}
}

@article{skodras2001jpeg,
  title={The JPEG 2000 still image compression standard},
  author={Skodras, Athanassios and Christopoulos, Charilaos and Ebrahimi, Touradj},
  journal={IEEE Signal processing magazine},
  volume={18},
  number={5},
  pages={36--58},
  year={2001},
  publisher={IEEE}
}

@incollection{salomon2002data,
  title={Data compression},
  author={Salomon, David},
  booktitle={Handbook of massive data sets},
  pages={245--309},
  year={2002},
  publisher={Springer}
}

@article{mildenberger2002introduction,
  title={Introduction to the DICOM standard},
  author={Mildenberger, Peter and Eichelberg, Marco and Martin, Eric},
  journal={European radiology},
  volume={12},
  number={4},
  pages={920--927},
  year={2002},
  publisher={Springer}
}

@article{rhodes2007locally,
  title={Locally optimal run-length compression applied to CT images},
  author={Rhodes, Michael L and Quinn, John F and Silvester, John},
  journal={IEEE transactions on medical imaging},
  volume={4},
  number={2},
  pages={84--90},
  year={2007},
  publisher={IEEE}
}

@article{wasserthal2023totalsegmentator,
  title={TotalSegmentator: robust segmentation of 104 anatomic structures in CT images},
  author={Wasserthal, Jakob and Breit, Hanns-Christian and Meyer, Manfred T and Pradella, Maurice and Hinck, Daniel and Sauter, Alexander W and Heye, Tobias and Boll, Daniel T and Cyriac, Joshy and Yang, Shan and others},
  journal={Radiology: Artificial Intelligence},
  volume={5},
  number={5},
  pages={e230024},
  year={2023},
  publisher={Radiological Society of North America}
}

@inproceedings{lin2014microsoft,
  title={Microsoft coco: Common objects in context},
  author={Lin, Tsung-Yi and Maire, Michael and Belongie, Serge and Hays, James and Perona, Pietro and Ramanan, Deva and Doll{\'a}r, Piotr and Zitnick, C Lawrence},
  booktitle={European conference on computer vision},
  pages={740--755},
  year={2014},
  organization={Springer}
}

@inproceedings{agustsson2019generative,
  title={Generative adversarial networks for extreme learned image compression},
  author={Agustsson, Eirikur and Tschannen, Michael and Mentzer, Fabian and Timofte, Radu and Gool, Luc Van},
  booktitle={Proceedings of the IEEE/CVF international conference on computer vision},
  pages={221--231},
  year={2019}
}

@inproceedings{iwai2021fidelity,
  title={Fidelity-controllable extreme image compression with generative adversarial networks},
  author={Iwai, Shoma and Miyazaki, Tomo and Sugaya, Yoshihiro and Omachi, Shinichiro},
  booktitle={2020 25th International Conference on Pattern Recognition (ICPR)},
  pages={8235--8242},
  year={2021},
  organization={IEEE}
}

@article{vikraman2024segmentation,
  title={Segmentation based medical image compression of brain magnetic resonance images using optimized convolutional neural network},
  author={Vikraman, Bindu Puthentharayil and Jabeena, A},
  journal={Multimedia Tools and Applications},
  volume={83},
  number={9},
  pages={26643--26661},
  year={2024},
  publisher={Springer}
}

\end{document}